%% file: main.tex
\documentclass[submission,copyright,creativecommons]{eptcs}
\providecommand{\event}{QPL 2022} 

\usepackage{appendix}

\usepackage{cite}
\usepackage{amsmath,amssymb,amsfonts,mathtools}
\usepackage{algorithmic}
\usepackage{graphicx}
\graphicspath{{figures/}{tikz/}}
\usepackage{wrapfig}
\usepackage{subcaption}
\usepackage[inline]{enumitem}
\usepackage{multicol}
\usepackage{vwcol}  
\usepackage{colortbl}
\usepackage{textcomp}
\usepackage{xcolor}
\def\BibTeX{{\rm B\kern-.05em{\sc i\kern-.025em b}\kern-.08em
    T\kern-.1667em\lower.7ex\hbox{E}\kern-.125emX}}

\usepackage[utf8]{inputenc} 
\usepackage{tabularx,booktabs}
\usepackage{braket}
\usepackage{tikz}
\usepackage{qcircuit}
\usepackage{tikz/tikzit}
\usepackage{multirow}
\input{tikz/sample.tikzstyles}
\usetikzlibrary{arrows.meta}

\DeclareGraphicsExtensions{.pdf,.jpeg,.png}

\input{reusable-functions}
\usepackage{hyperref}

\newboolean{review}
\reviewtrue
\reviewfalse

\ifreview
\paperwidth=\dimexpr \paperwidth + 12cm\relax
\oddsidemargin=\dimexpr\oddsidemargin + 6cm\relax
\evensidemargin=\dimexpr\evensidemargin + 6cm\relax
\marginparwidth=\dimexpr \marginparwidth + 6cm\relax
\else
\setuptodonotes{disable}
\fi

\title{Reducing 2-QuBit Gate Count for ZX-Calculus based Quantum Circuit Optimization}
\def\titlerunning{Reducing 2-QuBit Gate Count for ZX-Calculus based Quantum Circuit Optimization}
\def\authorrunning{Staudacher K., Guggemos T., Gehrke W., Grundner-Culemann S.}
\def\copyrightholders{Staudacher, Guggemos, Gehrke,\\\phantom{\copyright~}Grundner-Culemann} 

\input{content/authors}

\begin{document}

\maketitle

\input{content/abstract} 
\input{content/intro}
\input{content/background}

\input{content/related}
\input{content/heuristixz}
\input{content/neighbor-unfusion}
\input{content/evaluation}
\input{content/conclusion}

\input{content/acks}

\bibliographystyle{styles/eptcs}
\bibliography{Bib/paper,Bib/citavi}

\newpage

\begin{appendices}
\input{content/appendix-further-rules}
\newpage
\input{content/appendix-graph-theory}
\newpage
\input{content/appendix-zx-example}
\end{appendices}

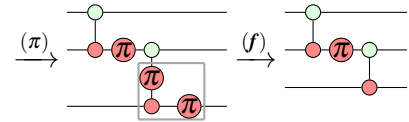
\end{document}

%% file: tikz/sample.tikzstyles
\tikzstyle{gate}=[shape=rectangle, text height=1.5ex, text depth=0.25ex, yshift=0.5mm, fill=white, draw=black, minimum height=5mm, yshift=-0.5mm, minimum width=5mm, font={\normalsize}, tikzit category=circuit]
\tikzstyle{big gate}=[shape=rectangle, text height=1.5ex, text depth=0.25ex, yshift=0.5mm, fill=white, draw=black, minimum height=10mm, yshift=-0.5mm, minimum width=5mm, font={\normalsize}, tikzit category=circuit]
\tikzstyle{Z dot}=[inner sep=0mm, minimum size=2mm, shape=circle, draw=black, fill={rgb,255: red,221; green,255; blue,221}, tikzit category=zx]
\tikzstyle{Z phase dot}=[minimum size=5mm, font={\footnotesize\boldmath}, shape=rectangle, rounded corners=2mm, inner sep=0.2mm, outer sep=-2mm, scale=0.8, tikzit shape=circle, draw=black, fill={rgb,255: red,221; green,255; blue,221}, tikzit draw=blue, tikzit category=zx]
\tikzstyle{X dot}=[Z dot, shape=circle, draw=black, fill={rgb,255: red,255; green,136; blue,136}, tikzit category=zx]
\tikzstyle{X phase dot}=[Z phase dot, tikzit shape=circle, tikzit draw=blue, fill={rgb,255: red,255; green,136; blue,136}, font={\footnotesize\boldmath}, tikzit category=zx]
\tikzstyle{hadamard}=[fill=yellow, draw=black, shape=rectangle, inner sep=0.6mm, minimum height=1.5mm, minimum width=1.5mm, tikzit category=zx]
\tikzstyle{paulibox}=[fill={rgb,255: red,221; green,221; blue,255}, draw=black, shape=rectangle, inner sep=0.6mm, minimum height=5mm, minimum width=5mm, font={\footnotesize}, text height=1.5ex, text depth=0.25ex, tikzit category=zx]
\tikzstyle{vertex}=[inner sep=0mm, minimum size=1mm, shape=circle, draw=black, fill=black, tikzit category=misc]
\tikzstyle{vertex set}=[inner sep=0mm, minimum size=1mm, shape=circle, draw=black, fill=white, font={\footnotesize\boldmath}, tikzit category=misc]
\tikzstyle{small black dot}=[fill=black, draw=black, shape=circle, inner sep=0pt, minimum width=1.2mm, tikzit category=circuit]
\tikzstyle{cnot ctrl}=[fill=black, draw=black, shape=circle, inner sep=0pt, minimum width=1.2mm, tikzit category=circuit]
\tikzstyle{cnot targ}=[fill=white, draw=white, shape=circle, tikzit category=circuit, label={center:$\oplus$}, inner sep=0pt, minimum width=2.1mm, tikzit fill={rgb,255: red,102; green,204; blue,255}, tikzit draw=black]
\tikzstyle{ket}=[fill=white, draw=black, shape=regular polygon, regular polygon sides=3, regular polygon rotate=-30, scale=0.7, inner sep=1pt, tikzit category=circuit, tikzit shape=rectangle, tikzit fill=green]
\tikzstyle{bra}=[fill=white, draw=black, shape=regular polygon, regular polygon sides=3, regular polygon rotate=30, scale=0.7, inner sep=1pt, tikzit category=circuit, tikzit shape=rectangle, tikzit fill=red]
\tikzstyle{scalar}=[shape=rectangle, text height=1.5ex, text depth=0.25ex, yshift=0.5mm, fill=white, draw=black, minimum height=5mm, yshift=-0.5mm, minimum width=5mm, font={\normalsize}]
\tikzstyle{clabel}=[fill=white, draw=none, shape=rectangle, tikzit fill={rgb,255: red,56; green,255; blue,242}, font={\footnotesize}, inner sep=1pt, tikzit category=labels]
\tikzstyle{empty diagram}=[draw={gray!40!white}, dashed, shape=rectangle, minimum width=1cm, minimum height=1cm, tikzit category=misc]

\tikzstyle{hadamard edge}=[-, dashed, dash pattern=on 2pt off 0.5pt, thick, draw={rgb,255: red,68; green,136; blue,255}]
\tikzstyle{box edge}=[-, dashed, dash pattern=on 2pt off 0.5pt, thick, draw={rgb,255: red,203; green,192; blue,225}]
\tikzstyle{brace edge}=[-, tikzit draw=blue, decorate, decoration={brace,amplitude=1mm,raise=-1mm}]
\tikzstyle{diredge}=[->]
\tikzstyle{double edge}=[-, double, shorten <=-1mm, shorten >=-1mm, double distance=2pt]
\tikzstyle{gray edge}=[-, {gray!70!white}, thick]
\tikzstyle{pointer edge}=[->, very thick, gray]
\tikzstyle{boldedge}=[-, line width=1.6pt, shorten <=-0.17mm, shorten >=-0.17mm]

%% file: reusable-functions.tex
\usepackage{todonotes}
\usepackage{easyReview}
\usepackage{xspace} 
\usepackage{pifont}
\usepackage{tcolorbox}
\usepackage{csquotes}

\newcommand{\content}[1]{}
\newcommand{\short}[1]{}

\newcolumntype{d}[1]{D{.}{.}{#1}}
\newcolumntype{L}[1]{>{\raggedright\let\newline\\\arraybackslash\hspace{0pt}}m{#1}}
\newcolumntype{C}[1]{>{\centering\let\newline\\\arraybackslash\hspace{0pt}}m{#1}}
\newcolumntype{R}[1]{>{\raggedleft\let\newline\\\arraybackslash\hspace{0pt}}m{#1}}

\newcommand{\myurl}[1]{{\footnotesize\texttt{\url{#1}}}}

\usepackage[framemethod=tikz]{mdframed}
\newmdtheoremenv[style=plain,backgroundcolor=gray!7]{definition}{Definition}

\newtheorem{proposition}{Proposition}[section]
\newcommand{\rfc}[2][0]{%
    \ifnum#1=0
      RFC\,#2\xspace
    \else
      RFC\,#2~\cite{RFC#2}\xspace
    \fi
}

\makeatletter%
\@ifclassloaded{IEEEtran}{
\usepackage[framemethod=tikz]{mdframed}
\newmdtheoremenv[style=plain,backgroundcolor=gray!5]{proposition}{Proposition}

}{}
\makeatother

%% file: content/authors.tex
\author{Korbinian Staudacher, Tobias Guggemos, Sophia Grundner-Culemann
	\institute{\textit{MNM-Team} \\
	Ludwig-Maximilians-Universität München\\
	Munich, Germany}
	\email{\{staudacher,guggemos,grundner-culemann\}@nm.ifi.lmu.de}
	\and
	Wolfgang Gehrke
	\institute{\textit{Research Institute CODE}\\
		Universität der Bundeswehr München
		Munich, Germany}
	\email{wolfgang.gehrke@unibw.de}
}

%% file: content/abstract.tex

\begin{abstract}
In the near term, programming quantum computers will remain severely limited by low quantum volumes.
Therefore, it is desirable to implement quantum circuits with the fewest resources possible.
For the common Clifford+T circuits, most research is focused on reducing the number of T gates, since they are an order of magnitude more expensive than Clifford gates in quantum error corrected encoding schemes.
However, this optimization sometimes leads to more 2-qubit gates, which, even though they are less expensive in terms of fault-tolerance, contribute significantly to the overall circuit cost.
Approaches based on the \textit{ZX-calculus} have recently gained some popularity in the field, but reduction of 2-qubit gates is not their focus.
In this work, we present an alternative for improving 2-qubit gate count of a quantum circuit with the ZX-calculus by using heuristics in ZX-diagram simplification.
Our approach maintains the good reduction of the T gate count provided by other strategies based on ZX-calculus, thus serving as an extension for other optimization algorithms.
Our results show that combining the available ZX-calculus-based optimizations with our algorithms can reduce the number of 2-qubit gates by as much as 40\,\% compared to current approaches using ZX-calculus.
Additionally, we improve the results of the best currently available optimization technique of Nam et.\,al\,\cite{nam2018automated} for some circuits  by up to 15\,\%.
\end{abstract}


%% file: content/intro.tex
\section{Introduction}
Many famous quantum algorithms, like Shor~\cite{Shor.1994}, HHL~\cite{Harrow.2009} or Grover~\cite{Grover.1996}, base upon techniques like Quantum Fourier Transformation, Quantum Phase Estimation or Amplification, respectively.
Although these algorithms provide significant (sometimes even exponential) speed-ups, current quantum chips can only execute toy problems, mostly due to the low gate fidelity. 
Even for problems that can be easily solved on a state-of-the-art desktop PC, those algorithms require tens of thousands of gates, and are therefore infeasible to run on near-term quantum devices.
However, applications in quantum simulation are supposed to achieve significant improvements in quantum chemistry, material sciences, or high-energy physics on near-term devices.
With variational algorithms (e.g., QAOA~\cite{Farhi.2014} or VQE~\cite{Peruzzo.2014}), real-world applications like optimization problems on real quantum chips may become feasible.
While the associated speed-up is unknown for many use cases, they require only few qubits and quantum gates to achieve promising results.
Quantum Machine Learning (QML) is such an example:
Here, the combination of clever encoding strategies, variational algorithms, and classical pre- and post-processing achieves high accurate classification rates with fewer qubits compared to classical bits.

Still, even algorithms with smaller circuits cannot be executed on current devices and gate optimization is a vibrant research topic.
While global optimization of arbitrary quantum circuits is generally QMA-hard~\cite{Janzing.2003}, different algorithms like quantum optimal control~\cite{Khaneja.2001} have been proposed to reduce the size of a quantum circuit.
In this context the so-called ZX-calculus~\cite{Coecke.2011} is considered a promising tool. 
It provides an abstract graphical language for describing quantum systems and can be seen as an alternative to the predominant description in the Hilbert space. 
We can transform any quantum circuit into a ZX-diagram equivalent, apply the rules of the ZX-calculus to simplify the diagram, and re-extract a quantum circuit from it.

\subsection*{Scope of this work}


Our work is based on optimizing circuits with ZX-calculus, where several optimization strategies have been proposed recently \cite{kissinger2020reducing,duncan2020graph,sivarajah2020t}.
Currently, these strategies yield very good results for pure \textit{Clifford} circuits, as well as for T gate elimination in \textit{Clifford+T} circuits, which is worthwhile for fault-tolerant quantum computers with error-corrected gates.
For such devices, the cost of a T gate is sometimes estimated to be up to a hundred times higher than the cost of a CNOT gate~\cite{OGorman.2017} (even though recent studies suggest lower rates~\cite{Litinski.2019}).

However, reducing 2-qubit gates is generally of interest for quantum hardware that is not error corrected (e.g., NISQ devices) or in which quantum states do not tend to interact easily, e.g., in Photonic Quantum computing~\cite{Barz.2015}.
A major drawback of the current ZX-calculus based strategies is that these gates in particular are not optimized very well; in fact, for many large Clifford+T circuits, the 2-qubit gate count even increases when using algorithms like the one in~\cite{duncan2020graph}.

We propose new optimization approaches especially for reducing 2-qubit gates.
To do so, we use heuristics for estimating the 2-qubit gate count in ZX-diagrams as cost functions for classical search algorithms like 
\begin{enumerate*}[label=\textbf{\Roman*.)}]
	\item random selection and
	\item greedy algorithm.
\end{enumerate*}
By combining them with existing optimization approaches, we maintain the T gate count reduction rate and improve the total gate count and the 2-qubit gate count for most given circuits. 
We evaluate the performance on circuits from the \textit{Tpar} benchmark~\cite{amy2014polynomial}.
We find that our optimizations can outperform existing ZX-based approaches and can additionally be used to further improve already optimized circuits.

%% file: content/background.tex

\section{Background}
\label{sec:background}
Throughout this paper, we use the notation from \cite{Nielsen.2013} for quantum gates; the most essential ones are detailed in \autoref{tab:overview-spiders} by name and matrix-, gate- and ZX-calculus-representation. 
Every unitary operation can be decomposed into a combination of CNOTs and single-qubit gates~\cite{Nielsen.2013}.
A well-studied example for a minimal gate set with which to approximate any unitary operation is the so-called \textit{Clifford+T set}, i.e., the gate set generated by $\{H,T,CNOT\}$.
That is why many optimization algorithms target circuits generated with the \textit{Clifford+T set}.
%
The \textit{Clifford} set generated by $\{H,S,CNOT\}$ is also well-known and useful for quantum circuit simulations on classical computers, but not every unitary operation can be represented with it.
For convenience, we abbreviate some gates in the \textit{Clifford+T} set, namely $X,Y,Z,S$, and $CZ$ (instead of writing, for example, $ Z = T\cdot T\cdot T\cdot T $).
%

\subsection{ZX-Calculus}
Since ZX-calculus and its optimization strategies rely on graph theory, we provide some background in \autoref{sec:background-graph}.
The ZX-calculus~\cite{coecke2018picturing,coecke2021kindergarden} is a graphical language for expressing linear maps on qubits as \textit{ZX-diagrams}. Relations in multi-qubit systems are often difficult to understand in Dirac notation, since the matrix size doubles with every qubit and the complex number space quickly becomes confusing. 
\begin{wrapfigure}{r}{0.28\textwidth}
	\vspace{-1.5em}
	\begin{equation}\label{eq:spiders}
		\begin{aligned}
			\input{greenSpider.tikz} \qquad& \input{redSpider.tikz}	\\
			\text{\textit{Z-Spider}} \qquad& \text{\textit{X-Spider}}
		\end{aligned}
	\end{equation}
\vspace{-2em}
\end{wrapfigure}

ZX-calculus provides a way to represent quantum circuits as 2-dimensional diagrams where nodes (\textit{spiders}) and edges (\textit{wires}) form an undirected graph.
In contrast to quantum circuits, the number of input- and output wires does not have to match, hence the resulting transformations are not necessarily unitary.
However, many important concepts in quantum mechanics follow very intuitively from this representation and we will briefly introduce the main principles.

\subsubsection{Representing Quantum Circuits}

\begin{wrapfigure}{r}{0.4\textwidth}
	\vspace{-2em}
	\begin{equation}\label{eq:zx-circuit}
		\begin{minipage}{0.5\linewidth}
			\Qcircuit @C=0.5em @R=.5em @!R {
				& \ctrl{1} & \qw & \qw  & \qw  & \qw \\
				& \targ & \gate{Z} & \ctrl{1} & \gate{Z} & \qw \\
				& \qw & \qw      & \targ    & \qw      & \qw
		}\end{minipage} ~ = ~ \input{simplethreeq_zx.tikz} \\
	\end{equation}
	\vspace{-2.5em}
\end{wrapfigure}

Any transformation on a single qubit can be described as a rotation around the $ X $ and $ Z $ axes.
Further, we can represent any quantum gate as a combination of $ X $- (red) and $ Z $-spiders (green) in ZX-diagrams (c.f.~\autoref{eq:spiders}), of which the most important are shown in \autoref{tab:overview-spiders}.
We call the wires on the left- and rightmost the \textit{input} and \textit{output} of the graph, respectively.
The three generators of the universal \textit{Clifford+T} set are constructed with the H-wire, the Z-spider with phase $\pi/4$, and a combination of an empty $ X $- and $ Z $-spiders (phase $ \alpha = 0 $) representing a CNOT.
In general, we can read a ZX-diagram in any direction since only the connectivity of the spiders matters, but for comparison with common quantum circuits it is convenient to read ZX-diagrams horizontally as shown in \autoref{eq:zx-circuit}.



\begin{table*}[t]
	\caption{Overview of important quantum gates and the respective ZX-Spiders.}
	\label{tab:overview-spiders}
	\centering
	\input{tabulars/overview-zx-spiders}
\end{table*}

\subsubsection{Basic Rules}
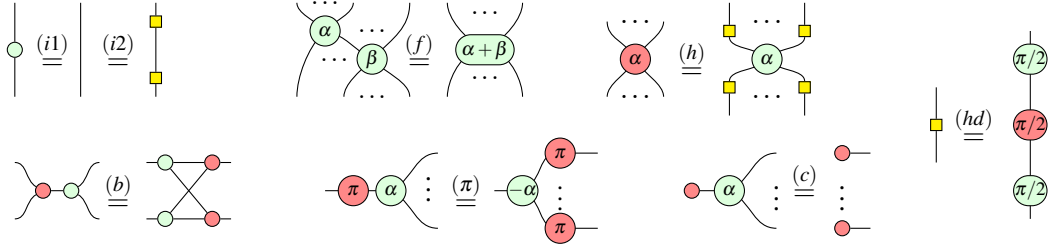
\begin{figure}[t]
	\begin{equation*}
	\input{all_rules.tikz}
	\end{equation*}
\caption{The important rules in ZX-Calculus that can be used for optimization are: Identity- (i1,i2), Fusion- (f), Hadamard- (h), Bialgebra- (b), Pi- ($ \pi $), Copy- (c) and Hadamard-Decomposition (hd) rule. Each holds for all $ \alpha,\beta \in [0,2\pi] $. Due to (h) and (i2), all rules hold with the colours interchanged.}
	\label{fig:all_rules}
\end{figure}

We introduce the most important transformation rules in the ZX language that are useful for optimization~\cite{Coecke.2011} in \autoref{fig:all_rules}.
All ZX-rules can be applied in both directions and also apply with inverted colors.

Any two \textit{Clifford} diagrams (i.e., diagrams only containing spiders with a \textit{Clifford} phase $ \alpha = k \cdot \frac{\pi}{2}; k \in \mathbb{Z} $) that represent the same linear map can be transformed into each other by some combination of those rules. 
Recent developments have introduced rule sets where this is also possible for \textit{Clifford+T} diagrams and for all ZX-diagrams~\cite{jeandel2018complete, vilmart2019near}.


%% file: tikz/greenSpider.tikz
\begin{tikzpicture}
	\begin{pgfonlayer}{nodelayer}
		\node [style=Z phase dot] (0) at (0, 0) {$\alpha$};
		\node [style=none] (1) at (1.25, 1.5) {};
		\node [style=none] (2) at (-1.25, 1.5) {};
		\node [style=none] (5) at (-1.25, -1.5) {};
		\node [style=none] (6) at (1.25, -1.5) {};
		\node [style=none] (7) at (0, -1) {$\ldots$};
		\node [style=none] (8) at (0, 1) {$\ldots$};
		\node [style=none] (9) at (0, 1.5) {$n$};
		\node [style=none] (10) at (0, -1.5) {$m$};
	\end{pgfonlayer}
	\begin{pgfonlayer}{edgelayer}
		\draw [in=-90, out=60, looseness=0.75] (0) to (1.center);
		\draw [in=-90, out=120] (0) to (2.center);
		\draw [in=90, out=-120, looseness=0.75] (0) to (5.center);
		\draw [in=90, out=-60, looseness=0.75] (0) to (6.center);
	\end{pgfonlayer}
\end{tikzpicture}

%% file: tikz/redSpider.tikz
\begin{tikzpicture}
	\begin{pgfonlayer}{nodelayer}
		\node [style=X phase dot] (0) at (0, 0) {$\alpha$};
		\node [style=none] (1) at (1.25, 1.5) {};
		\node [style=none] (2) at (-1.25, 1.5) {};
		\node [style=none] (5) at (-1.25, -1.5) {};
		\node [style=none] (6) at (1.25, -1.5) {};
		\node [style=none] (7) at (0, -1) {$\ldots$};
		\node [style=none] (8) at (0, 1) {$\ldots$};
		\node [style=none] (9) at (0, 1.5) {$n$};
		\node [style=none] (10) at (0, -1.5) {$m$};
	\end{pgfonlayer}
	\begin{pgfonlayer}{edgelayer}
		\draw [in=-90, out=60, looseness=0.75] (0) to (1.center);
		\draw [in=-90, out=120] (0) to (2.center);
		\draw [in=90, out=-120, looseness=0.75] (0) to (5.center);
		\draw [in=90, out=-60, looseness=0.75] (0) to (6.center);
	\end{pgfonlayer}
\end{tikzpicture}

%% file: tikz/simplethreeq_zx.tikz
\begin{tikzpicture}
	\begin{pgfonlayer}{nodelayer}
		\node [style=none] (0) at (-4, 1.5) {};
		\node [style=none] (1) at (-4, 0.1) {};
		\node [style=none] (2) at (-4, -1.25) {};
		\node [style=Z dot] (3) at (-3.25, 1.5) {};
		\node [style=Z phase dot] (4) at (-2.25, 0.1) {$\pi$};
		\node [style=Z dot] (5) at (-1.25, 0.1) {};
		\node [style=Z phase dot] (6) at (-0.25, 0.1) {$\pi$};
		\node [style=X dot] (7) at (-3.25, 0.1) {};
		\node [style=X dot] (8) at (-1.25, -1.25) {};
		\node [style=none] (9) at (0.5, 1.5) {};
		\node [style=none] (10) at (0.5, 0.1) {};
		\node [style=none] (11) at (0.5, -1.25) {};
	\end{pgfonlayer}
	\begin{pgfonlayer}{edgelayer}
		\draw (0.center) to (3);
		\draw (7) to (1.center);
		\draw (2.center) to (8);
		\draw (11.center) to (8);
		\draw (6) to (10.center);
		\draw (6) to (5);
		\draw (5) to (4);
		\draw (4) to (7);
		\draw (5) to (8);
		\draw (3) to (7);
		\draw (3) to (9.center);
	\end{pgfonlayer}
\end{tikzpicture}

%% file: tabulars/overview-zx-spiders.tex
\setlength\arraycolsep{1.5pt}
\resizebox{\textwidth}{!}{
\begin{tabular}{C{1.25cm}|C{1cm}C{1.3cm}C{1.6cm}C{1.2cm}C{1cm}C{3cm}C{1.8cm}C{2.5cm}}
	\textbf{Name}    & \textbf{Identity} & \textbf{Z} & \textbf{Z-Phase} & \textbf{T} & \textbf{X} & \textbf{X-Phase} & \textbf{H} & \textbf{CNOT}\\[.5em]\hline
	\textbf{Matrix}  & $ \begin{pmatrix}1 & 0 \\ 0 & 1\end{pmatrix} $ 
	& $ \begin{pmatrix}1 & 0 \\ 0 & -1\end{pmatrix} $ 
	& \quad$ \begin{pmatrix}1 & 0 \\ 0 & e^{i\alpha}\end{pmatrix} $\quad
	& $ \begin{pmatrix}1 & 0 \\ 0 & e^{\frac{i\pi}{4}}\end{pmatrix} $  	
	& $ \begin{pmatrix}0 & 1 \\ 1 & 0\end{pmatrix} $  
	& $\frac{1}{2} \begin{pmatrix}1 + e^{i\alpha} & 1 - e^{i\alpha} \\ 1 - e^{i\alpha} & 1 + e^{i\alpha}\end{pmatrix} $  
	& $ \frac{1}{\sqrt{2}}\begin{pmatrix}1 & 1 \\ 1 & -1\end{pmatrix} $ 
	& $ \begin{pmatrix} 1 & 0 & 0 & 0 \\ 0 & 1 & 0 & 0 \\ 0 & 0 & 0 & 1 \\ 0 & 0 & 1 & 0 \end{pmatrix} $ \\[.5em]
	\textbf{Gate}    & $ \Qcircuit @C=.5em @R=.75em {& \gate{I} & \qw} $ 
	& $ \Qcircuit @C=.5em @R=.75em {& \gate{Z} & \qw} $ 
	& $ \Qcircuit @C=.5em @R=.75em {& \gate{R_z(\alpha)} & \qw} $
	& $ \Qcircuit @C=.5em @R=.75em {& \gate{T} & \qw} $
	& $ \Qcircuit @C=.5em @R=.75em {& \gate{X} & \qw} $
	& $ \Qcircuit @C=.5em @R=.75em {& \gate{R_x(\alpha)} & \qw} $
	& $ \Qcircuit @C=.5em @R=.75em {& \gate{H} & \qw} $
	& $ \Qcircuit @C=.5em @R=1em {& \ctrl{1} & \qw \\ & \targ & \qw} $ \\[-1em]
	\textbf{Spider/ Wire}  & \ctikzfig{simple_wire} & \ctikzfig{PauliZ}  & \ctikzfig{ZPhaseSpider} & \ctikzfig{TGate}  & \ctikzfig{PauliX} & \ctikzfig{XPhaseSpider} & \ctikzfig{hadamardSingle} & \ctikzfig{CNOT}				 
\end{tabular}
}

%% file: tikz/all_rules.tikz
\begin{tikzpicture}
	\begin{pgfonlayer}{nodelayer}
		\node [style=none] (0) at (8, -2.75) {$\overset{(c)}{=}$};
		\node [style=none] (1) at (7.25, -2) {};
		\node [style=none] (2) at (7.25, -3) {\vdots};
		\node [style=none] (3) at (7.25, -4) {};
		\node [style=Z phase dot] (4) at (6, -3) {$\alpha$};
		\node [style=X dot] (5) at (5, -3) {};
		\node [style=none] (6) at (9.75, -2) {};
		\node [style=none] (7) at (9, -3) {\vdots};
		\node [style=none] (8) at (9.75, -4) {};
		\node [style=X dot] (9) at (9, -2) {};
		\node [style=X dot] (10) at (9, -4) {};
		\node [style=none] (11) at (-2.25, 0.75) {$\overset{(f)}{=}$};
		\node [style=Z phase dot] (12) at (-3.5, 0.5) {$\beta$};
		\node [style=none] (13) at (-2.5, -0.5) {};
		\node [style=none] (14) at (-4.25, -0.5) {};
		\node [style=Z phase dot] (15) at (-4.75, 1.25) {$\alpha$};
		\node [style=none] (16) at (-4, 2) {};
		\node [style=none] (17) at (-5.5, 2) {};
		\node [style=none] (18) at (-5.5, -0.5) {};
		\node [style=Z phase dot] (19) at (-0.5, 0.75) {~$\alpha+\beta$~~};
		\node [style=none] (20) at (-1.5, 2) {};
		\node [style=none] (21) at (-1.5, -0.5) {};
		\node [style=none] (22) at (0.5, -0.5) {};
		\node [style=none] (23) at (0.5, 2) {};
		\node [style=none] (24) at (-4.5, 0.5) {\small$\ldots$};
		\node [style=none] (25) at (-4.75, 2) {\small$\ldots$};
		\node [style=none] (26) at (-0.5, 1.75) {\small$\ldots$};
		\node [style=none] (27) at (-0.5, 0) {\small$\ldots$};
		\node [style=none] (28) at (-3.5, -0.5) {\small$\ldots$};
		\node [style=none] (29) at (-2.5, 2) {};
		\node [style=none] (30) at (-3.5, 1.25) {\small$\ldots$};
		\node [style=none] (31) at (-10.25, 0.75) {$\overset{(i2)}{=}$};
		\node [style=none] (32) at (-12, 0.75) {$\overset{(i1)}{=}$};
		\node [style=none] (33) at (-11.25, 2) {};
		\node [style=none] (34) at (-11.25, -0.5) {};
		\node [style=hadamard] (35) at (-9.25, 1.5) {};
		\node [style=none] (36) at (-9.25, 2) {};
		\node [style=none] (37) at (-9.25, -0.5) {};
		\node [style=none] (38) at (-13, 2) {};
		\node [style=Z dot] (39) at (-13, 0.75) {};
		\node [style=none] (40) at (-13, -0.5) {};
		\node [style=hadamard] (41) at (-9.25, 0) {};
		\node [style=none] (42) at (-10.25, -3) {$\overset{(b)}{=}$};
		\node [style=none] (43) at (-10.75, -2.25) {};
		\node [style=none] (44) at (-10.75, -3.75) {};
		\node [style=Z dot] (45) at (-11.5, -3) {};
		\node [style=X dot] (46) at (-12.25, -3) {};
		\node [style=none] (47) at (-13, -2.25) {};
		\node [style=none] (48) at (-13, -3.75) {};
		\node [style=none] (49) at (-9.5, -2.25) {};
		\node [style=none] (50) at (-9.5, -3.75) {};
		\node [style=Z dot] (51) at (-9, -2.25) {};
		\node [style=Z dot] (52) at (-9, -3.75) {};
		\node [style=X dot] (53) at (-7.75, -3.75) {};
		\node [style=X dot] (54) at (-7.75, -2.25) {};
		\node [style=none] (55) at (-7.25, -2.25) {};
		\node [style=none] (56) at (-7.25, -3.75) {};
		\node [style=none] (57) at (-1, -3) {$\overset{(\pi)}{=}$};
		\node [style=none] (58) at (-1.75, -2) {};
		\node [style=none] (59) at (-2, -2.75) {\vdots};
		\node [style=none] (60) at (-1.75, -4) {};
		\node [style=Z phase dot] (61) at (-3, -3) {$\alpha$};
		\node [style=X phase dot] (62) at (-4, -3) {$\pi$};
		\node [style=none] (63) at (-4.75, -3) {};
		\node [style=none] (64) at (-0.25, -3) {};
		\node [style=Z phase dot] (65) at (0.5, -3) {$-\alpha$};
		\node [style=none] (66) at (2.5, -2) {};
		\node [style=none] (67) at (1.5, -3) {\small\vdots};
		\node [style=none] (68) at (2.5, -4) {};
		\node [style=X phase dot] (69) at (1.5, -2) {$\pi$};
		\node [style=X phase dot] (70) at (1.5, -4) {$\pi$};
		\node [style=X phase dot] (71) at (3.5, 0.5) {$\alpha$};
		\node [style=none] (72) at (4.25, 1.5) {};
		\node [style=none] (73) at (2.75, 1.5) {};
		\node [style=none] (74) at (2.75, -0.5) {};
		\node [style=none] (75) at (4.25, -0.5) {};
		\node [style=none] (76) at (3.5, -0.5) {$\ldots$};
		\node [style=none] (77) at (3.5, 1.5) {$\ldots$};
		\node [style=none] (78) at (5, 0.5) {$\overset{(h)}{=}$};
		\node [style=Z phase dot] (79) at (7, 0.5) {$\alpha$};
		\node [style=none] (80) at (8, 2) {};
		\node [style=none] (81) at (6, 2) {};
		\node [style=none] (82) at (6, -1) {};
		\node [style=none] (83) at (8, -1) {};
		\node [style=none] (84) at (7, -0.5) {$\ldots$};
		\node [style=none] (85) at (7, 1.5) {$\ldots$};
		\node [style=hadamard] (86) at (6, 1.25) {};
		\node [style=hadamard] (87) at (8, 1.25) {};
		\node [style=hadamard] (88) at (8, -0.25) {};
		\node [style=hadamard] (89) at (6, -0.25) {};
		\node [style=hadamard] (90) at (11.5, -1.25) {};
		\node [style=none] (91) at (11.5, -0.25) {};
		\node [style=none] (92) at (11.5, -2.25) {};
		\node [style=none] (93) at (14, 1.25) {};
		\node [style=none] (95) at (14, -3.75) {};
		\node [style=X phase dot] (96) at (14, -1.25) {$\pi/2$};
		\node [style=Z phase dot] (97) at (14, 0.5) {$\pi/2$};
		\node [style=Z phase dot] (98) at (14, -3) {$\pi/2$};
		\node [style=none] (99) at (12.5, -1.25) {$\overset{(hd)}{=}$};
	\end{pgfonlayer}
	\begin{pgfonlayer}{edgelayer}
		\draw (5) to (4);
		\draw [in=-180, out=30] (4) to (1.center);
		\draw [in=-180, out=-30] (4) to (3.center);
		\draw (9) to (6.center);
		\draw (10) to (8.center);
		\draw [in=90, out=-45, looseness=0.75] (12) to (13.center);
		\draw [in=90, out=-135, looseness=0.75] (12) to (14.center);
		\draw [in=-135, out=90, looseness=0.75] (18.center) to (15);
		\draw [in=-90, out=135, looseness=0.75] (15) to (17.center);
		\draw [in=-90, out=45, looseness=0.75] (15) to (16.center);
		\draw [in=150, out=-15] (15) to (12);
		\draw [in=-90, out=30, looseness=0.75] (19) to (23.center);
		\draw [in=90, out=-15, looseness=0.75] (19) to (22.center);
		\draw [in=90, out=-165, looseness=0.75] (19) to (21.center);
		\draw [in=-90, out=150, looseness=0.75] (19) to (20.center);
		\draw [in=-90, out=45] (12) to (29.center);
		\draw (38.center) to (39);
		\draw (39) to (40.center);
		\draw (33.center) to (34.center);
		\draw (36.center) to (35);
		\draw (35) to (41);
		\draw (41) to (37.center);
		\draw [in=150, out=0] (47.center) to (46);
		\draw [in=0, out=-150] (46) to (48.center);
		\draw (45) to (46);
		\draw [in=180, out=30] (45) to (43.center);
		\draw [in=-180, out=-30] (45) to (44.center);
		\draw (50.center) to (52);
		\draw (52) to (54);
		\draw (49.center) to (51);
		\draw (51) to (53);
		\draw (53) to (56.center);
		\draw (54) to (55.center);
		\draw (51) to (54);
		\draw (52) to (53);
		\draw (63.center) to (62);
		\draw (62) to (61);
		\draw [in=-180, out=30] (61) to (58.center);
		\draw [in=-180, out=-30] (61) to (60.center);
		\draw (64.center) to (65);
		\draw [in=-180, out=30] (65) to (69);
		\draw (69) to (66.center);
		\draw [in=180, out=-30] (65) to (70);
		\draw (70) to (68.center);
		\draw [in=-90, out=60, looseness=0.75] (71) to (72.center);
		\draw [in=-90, out=120] (71) to (73.center);
		\draw [in=90, out=-120, looseness=0.75] (71) to (74.center);
		\draw [in=90, out=-60, looseness=0.75] (71) to (75.center);
		\draw [in=90, out=-135, looseness=0.75] (79) to (89);
		\draw [in=-90, out=135, looseness=0.75] (79) to (86);
		\draw [in=-90, out=45, looseness=0.75] (79) to (87);
		\draw (87) to (80.center);
		\draw (86) to (81.center);
		\draw (89) to (82.center);
		\draw [in=90, out=-90] (88) to (83.center);
		\draw [in=-45, out=90, looseness=0.75] (88) to (79);
		\draw (91.center) to (90);
		\draw (90) to (92.center);
		\draw (93.center) to (97);
		\draw (97) to (96);
		\draw (96) to (98);
		\draw (95.center) to (98);
	\end{pgfonlayer}
\end{tikzpicture}

%% file: content/related.tex

\section{Circuit optimization with ZX-calculus} \label{sec:related}
With the rules of ZX-calculus, the optimization of quantum circuits becomes a \textit{simplification} problem on the ZX-diagram. By simplification we mean reducing the total number of either spiders or wires in a diagram in order to obtain a smaller diagram.
The general process is as follows:
\begin{enumerate}[label=\arabic*.)]
	\item Transform the circuit to a ZX-diagram
	\item (optional:) transform to a \textit{graph-like} diagram, i.e.:\\[-2em]
	\begin{itemize}[leftmargin=.5em]
		\begin{multicols}{2}
			\item All spiders are Z-spiders.
			\item All connections are Hadamard wires.
			\item There are no loops. 
			\item Inputs and outputs are the only non-Hadamard wires and are connected to at most one spider. Every spider has at most one input and one output.
		\end{multicols}
	\end{itemize}
	\item Simplify the diagram using ZX-rules.
	\item Extract a quantum circuit out of the ZX-diagram.
\end{enumerate}
This allows powerful optimization of circuits, which are not obvious at a first glance (we provide an intuitive example in \autoref{app:zx-example}).

\subsection{Diagram simplification}
The presented ZX-rules allow many degrees of freedom, hence, simplification is still a difficult problem. 
The term ``simplification of diagrams'' has to be taken with a grain of salt since decreasing the number of spiders in a diagram can also lead to more complex extracted circuits.
Since rules can be applied in both directions it is important to find \textit{terminating} algorithms for diagram simplification. 
A common approach has been to only use ZX-rules which \textit{decrease} the total number of spiders in a diagram with every application, thus ensuring termination.  
We present some common approaches, many of which are implemented in the PyZX-library~\cite{kissinger2020Pyzx}.

\subsubsection{Clifford spider simplification}
The core of most strategies are two rules from graph theory -- namely \textit{local complementation} and \textit{pivoting}~-- which work on diagrams that are \textit{graph-like}. 
Both rules allow the elimination of interior Clifford spiders (phase $0,\pi/2,\pi,$ or $-\pi/2$; not connected to an input or output) from ZX-diagrams.

\paragraph{Local complementation ($ lc $)}

\begin{wrapfigure}{r}{0.32\textwidth}
	\vspace{-1.25em}
	\begin{equation}\label{eq:lcompRule}
		\input{lcompRuleExample.tikz}
	\end{equation}
\vspace{-1.75em}
\end{wrapfigure}
In ZX-calculus, local complementation from \autoref{sec:background-graph-lc} is applicable on graph-like diagrams.
If the spider $ a $ in $ G \star a $ has a phase of $\pm\pi/2$, the phase is subtracted from those of the neighboring spiders and the spider is eliminated for simplification of the diagram as shown in \autoref{eq:lcompRule}.

\paragraph{Pivoting ($ p $)}
Similarly we can eliminate a pair of spiders $uv$ with phase $0$ or $\pi$ by applying a graph-theoretic pivot $G \land uv$ (c.f. \autoref{sec:background-graph-pv}) on the diagram as in the following example ($ j,k \in \mathbb{Z} $):
\begin{equation}\label{eq:pivotRule}
	\input{pivotRuleExample.tikz}
\end{equation}

\subsubsection{Clifford simplification algorithm}
These rules allow constructing an algorithm for graph-like diagrams which removes most interior Clifford spiders~\cite{duncan2020graph}. The procedure is as follows: 
\begin{enumerate}
	\item Eliminate empty spiders with two wires using the identity rule and subsequently fuse the adjacent spiders in order to maintain a graph-like diagram.
	\item Apply local complementation on every spider of phase $\pm \pi/2$ and pivoting on every pair of connected spiders of phase $0$ or $\pi$ as often as possible.
	\item If step 2 modified the diagram, start again with step 1, else stop the iteration.
\end{enumerate}
That allows us to remove every interior spider with phase $\pm \pi/2$ and every pair of connected spiders with phase $0$ or $\pi$. 
However, after simplification some interior Clifford spiders with phase $0$ or $\pi$ may remain.

\subsubsection{Phase gadget simplification (p2)}

\begin{wrapfigure}{r}{0.2\textwidth}
	\vspace{-2.4em}
	\begin{equation}\label{eq:pivot_phase_gadget}
		\input{phase_gadget1.tikz}
	\end{equation}
	\vspace{-1.5em}
\end{wrapfigure}
We can use phase gadgets to apply pivoting on a pair of spiders where one spider has a non-Clifford phase ($ \neq 0,\pi/2,\pi,-\pi/2$). 
A phase gadget as defined in~\cite{kissinger2020reducing} is a parameterized spider with only one wire connected via Hadamard edge to a phaseless spider as in \autoref{eq:pivot_phase_gadget}.

We can modify pivoting (\autoref{eq:pivotRule}) to exchange the spider with a non-Clifford phase to a phase gadget: 
\begin{equation}\label{eq:pivotGadgetExample}
	\input{pivotGadgetExample.tikz}	
\end{equation}
With the additional rules in \autoref{app:further-rules}, we can eliminate every interior Clifford spider in a diagram~\cite{kissinger2020reducing}.

\subsection{Circuit extraction}\label{sec:related-extraction}
Extracting a quantum circuit from a simplified diagram can be challenging, since spiders with an arbitrary number of wires have no direct gate representation~\cite{backens2021there}.
The most general circuit extraction routine makes use of so called ``flow properties'' originating in measurement-based quantum computing (MBQC). 
Graph-like ZX-diagrams can be seen as an extension of MBQC graph states where the phases of spiders represent measurements in either the XY, XZ or YZ plane of the Bloch sphere:
\begin{equation}\label{measurement_planes}
	\renewcommand{\arraystretch}{1.25}
	\begin{tabular}{l || c | c | c}
		meas. plane & XY & XZ & YZ \\ \hline
		meas. effect & \input{mbqc_xy.tikz} & \input{mbqc_xz.tikz} = \input{mbqc_xz-graphlike.tikz} & \input{mbqc_yz.tikz} = \input{mbqc_yz-graphlike.tikz}  \\
	\end{tabular}
	\renewcommand{\arraystretch}{1} 
\end{equation}
Diagrams simplified with the Clifford simplification algorithm from~\cite{duncan2020graph} only contain spiders in the XY measurement plane and preserve a flow property called \textit{focused generalized flow (gflow)}. 
Those diagrams can be extracted by converting every spider with phase $\alpha$ to an $R_z\left(\alpha\right)$ gate and a Hadamard wire as either a $H$ or $CZ$ gate or a combination of $CNOT$ gates. 
As an example consider the diagrams from \autoref{eq:pivotRule}. Extracting the diagram on the left hand side yields the following circuit:
\begin{equation}\label{eq:extracted_circuit_unoptimized}\small
	\Qcircuit @C=.25em @R=.25em @!R {
		 & \gate{R_Z(\alpha_1)} & \gate{H} & \ctrl{1}  & \qw  & \qw & \ctrl{1} & \gate{R_Z(j\pi)} & \gate{H} & \ctrl{1} & \qw & \qw & \ctrl{1} & \gate{R_Z(k\pi)} & \gate{H} & \gate{R_Z(\gamma_1)} & \qw \\
		 & \qw & \qw & \control\qw & \gate{R_Z(\alpha_2)} & \gate{H} & \control\qw & \qw & \qw & \control\qw & \gate{R_Z(\beta_1)} & \gate{H} & \control\qw & \gate{R_Z(\gamma_2)} & \qw & \qw & \qw \\
	}
\end{equation}
whereas extracting the diagram after rule application yields the equivalent smaller circuit:
\begin{equation}\label{eq:extracted_circuit_optimized}\small
	\Qcircuit @C=.5em @R=.5em @!R {
		 & \qw & \qw & \ctrl{1}  & \gate{R_Z(\alpha_1')}  & \gate{H} & \gate{R_Z(\beta_1')} & \gate{H} & \ctrl{1} & \gate{R_Z(\gamma_1')} & \qw \\
		 & \gate{R_Z(\alpha_2')} & \gate{H} & \control\qw & \qw & \qw & \qw & \qw & \targ & \gate{R_Z(\gamma_2')} & \qw \\
	}
\end{equation}
However, ZX-diagrams simplified with \autoref{eq:pivotGadgetExample} may contain spiders in $XZ$ and $YZ$ plane as well. 
While it has been shown that those diagrams still preserve \textit{generalized flow (gflow)}, the circuit extraction routine has to convert those spiders back into $XY$ spiders using either pivoting $\left(YZ\right)$ or a combination of local complementation and pivoting $\left(XZ\right)$ before extracting the diagram~\cite{backens2021there}. 
An algorithm to efficiently extract diagrams that do not admit the gflow property is yet to be discovered; however, recent findings suggest that such an algorithm may not exist for general ZX-diagrams~\cite{de2022circuit}.
Hence, even though the diagram may represent a unitary matrix, we cannot extract a quantum circuit from the diagram efficiently.

%% file: tikz/lcompRuleExample.tikz
\begin{tikzpicture}
	\begin{pgfonlayer}{nodelayer}
		\node [style=Z phase dot] (0) at (-1.5, 0.75) {$\alpha_1$};
		\node [style=Z phase dot] (1) at (-3.5, 0.75) {$\pm\frac{\pi}{2}$};
		\node [style=Z phase dot] (2) at (-3.5, -1) {$\alpha_2$};
		\node [style=Z phase dot] (3) at (-1.5, -1) {$\alpha_3$};
		\node [style=none] (18) at (0, 0) {$=$};
		\node [style=Z phase dot] (19) at (3.5, 0.75) {$\alpha_1\mp\frac{\pi}{2}$};
		\node [style=Z phase dot] (21) at (1.25, -1) {$\alpha_2\mp\frac{\pi}{2}$};
		\node [style=Z phase dot] (22) at (3.5, -1) {$\alpha_3\mp\frac{\pi}{2}$};
		\node [style=none] (23) at (-3.5, 1.5) {$a$};
		\node [style=none] (24) at (0, 0.75) {$(lc)$};
		\node [style=none] (25) at (-4, 1.75) {};
		\node [style=none] (26) at (-4, 0.25) {};
		\node [style=none] (27) at (-3, 1.75) {};
		\node [style=none] (28) at (-3, 0.25) {};
	\end{pgfonlayer}
	\begin{pgfonlayer}{edgelayer}
		\draw [style=hadamard edge] (1) to (0);
		\draw [style=hadamard edge] (1) to (2);
		\draw [style=hadamard edge] (1) to (3);
		\draw [style=hadamard edge] (21) to (19);
		\draw [style=hadamard edge] (2) to (3);
		\draw [style=hadamard edge] (19) to (22);
		\draw [style=gray edge] (25.center) to (27.center);
		\draw [style=gray edge] (27.center) to (28.center);
		\draw [style=gray edge] (28.center) to (26.center);
		\draw [style=gray edge] (26.center) to (25.center);
	\end{pgfonlayer}
\end{tikzpicture}

%% file: tikz/pivotRuleExample.tikz
\begin{tikzpicture}
	\begin{pgfonlayer}{nodelayer}
		\node [style=Z phase dot] (0) at (-6.5, -0.75) {$\alpha_2$};
		\node [style=Z phase dot] (1) at (-6.5, 1) {$\alpha_1$};
		\node [style=Z phase dot] (2) at (-5.25, 1) {$j\pi$};
		\node [style=Z phase dot] (3) at (-2.75, 1) {$k\pi$};
		\node [style=Z phase dot] (4) at (-1.5, 1) {$\gamma_1$};
		\node [style=Z phase dot] (5) at (-4, -0.75) {$\beta_1$};
		\node [style=Z phase dot] (6) at (-1.5, -0.75) {$\gamma_2$};
		\node [style=none] (7) at (0, 0.25) {$\overset{(p)}{=}$};
		\node [style=Z phase dot] (8) at (1.75, 1) {$\alpha_1'$};
		\node [style=Z phase dot] (9) at (1.75, -0.75) {$\alpha_2'$};
		\node [style=Z phase dot] (12) at (6.25, -0.75) {$\gamma_2'$};
		\node [style=Z phase dot] (13) at (4, 2) {$\beta_1'$};
		\node [style=Z phase dot] (14) at (6.25, 1) {$\gamma_1'$};
		\node [style=none] (15) at (9.5, 1.5) {$\alpha_i'=\alpha_i+k\pi$};
		\node [style=none] (16) at (-5.25, 1.75) {$u$};
		\node [style=none] (17) at (-2.75, 1.75) {$v$};
		\node [style=none] (18) at (11, 0.5) {$\beta_i'=\beta_i+(j+k+1)\pi$};
		\node [style=none] (19) at (9.5, -0.5) {$\gamma_i'=\gamma_i+j\pi$};
		\node [style=none] (21) at (-5.75, 2) {};
		\node [style=none] (22) at (-5.75, 0.5) {};
		\node [style=none] (23) at (-2.25, 2) {};
		\node [style=none] (24) at (-2.25, 0.5) {};
		\node [style=none] (25) at (-7.25, 1) {};
		\node [style=none] (26) at (-7.25, -0.75) {};
		\node [style=none] (27) at (-0.75, 1) {};
		\node [style=none] (28) at (-0.75, -0.75) {};
		\node [style=none] (29) at (1, 1) {};
		\node [style=none] (30) at (1, -0.75) {};
		\node [style=none] (31) at (7, 1) {};
		\node [style=none] (32) at (7, -0.75) {};
	\end{pgfonlayer}
	\begin{pgfonlayer}{edgelayer}
		\draw [style=hadamard edge] (3) to (4);
		\draw [style=hadamard edge] (3) to (6);
		\draw [style=hadamard edge] (3) to (5);
		\draw [style=hadamard edge] (2) to (5);
		\draw [style=hadamard edge] (2) to (3);
		\draw [style=hadamard edge] (2) to (0);
		\draw [style=hadamard edge] (2) to (1);
		\draw [style=hadamard edge] (6) to (5);
		\draw [style=hadamard edge] (0) to (5);
		\draw [style=hadamard edge] (8) to (14);
		\draw [style=hadamard edge] (14) to (13);
		\draw [style=hadamard edge] (13) to (8);
		\draw [style=hadamard edge] (8) to (12);
		\draw [style=hadamard edge] (12) to (9);
		\draw [style=hadamard edge] (9) to (14);
		\draw [style=gray edge] (21.center) to (23.center);
		\draw [style=gray edge] (23.center) to (24.center);
		\draw [style=gray edge] (24.center) to (22.center);
		\draw [style=gray edge] (22.center) to (21.center);
		\draw (25.center) to (1);
		\draw (0) to (26.center);
		\draw (27.center) to (4);
		\draw (28.center) to (6);
		\draw (30.center) to (9);
		\draw (8) to (29.center);
		\draw (14) to (31.center);
		\draw (32.center) to (12);
	\end{pgfonlayer}
\end{tikzpicture}

%% file: tikz/phase_gadget1.tikz
\begin{tikzpicture}
	\begin{pgfonlayer}{nodelayer}
		\node [style=Z phase dot] (0) at (-1.5, 0) {$\alpha$};
		\node [style=Z dot] (1) at (0, 0) {};
		\node [style=none] (2) at (1.75, 1) {};
		\node [style=none] (3) at (1.75, -1) {};
		\node [style=none] (4) at (1.25, 0) {$\ldots$};
		\node [style=none] (5) at (-2.25, 1) {};
		\node [style=none] (6) at (1, 1) {};
		\node [style=none] (7) at (1, -1) {};
		\node [style=none] (8) at (-2.25, -1) {};
		\node [style=none] (10) at (-0.25, 1.75) {\textit{phase gadget}};
	\end{pgfonlayer}
	\begin{pgfonlayer}{edgelayer}
		\draw [style=hadamard edge] (0) to (1);
		\draw [in=-180, out=45, looseness=0.75] (1) to (2.center);
		\draw [in=-180, out=-45, looseness=0.75] (1) to (3.center);
		\draw [style=box edge] (5.center) to (6.center);
		\draw [style=box edge] (6.center) to (7.center);
		\draw [style=box edge] (7.center) to (8.center);
		\draw [style=box edge] (8.center) to (5.center);
	\end{pgfonlayer}
\end{tikzpicture}

%% file: tikz/pivotGadgetExample.tikz
\begin{tikzpicture}
	\begin{pgfonlayer}{nodelayer}
		\node [style=Z phase dot] (0) at (-7, -0.5) {$\alpha_2$};
		\node [style=Z phase dot] (1) at (-7, 1) {$\alpha_1$};
		\node [style=Z phase dot] (2) at (-5.75, 1) {$j\pi$};
		\node [style=Z phase dot] (3) at (-3.25, 1) {$\sigma$};
		\node [style=Z phase dot] (4) at (-2, 1) {$\gamma_1$};
		\node [style=Z phase dot] (5) at (-4.5, -0.5) {$\beta_1$};
		\node [style=Z phase dot] (6) at (-2, -0.5) {$\gamma_2$};
		\node [style=none] (7) at (-0.25, 0.5) {$\overset{(p2)}{=}$};
		\node [style=none] (16) at (-5.75, 1.75) {$u$};
		\node [style=none] (17) at (-3.25, 1.75) {$v$};
		\node [style=none] (18) at (9.675, 0.75) {$\beta_i'=\beta_i+(j+1)\pi$};
		\node [style=none] (19) at (8.75, -0.25) {$\gamma_i'=\gamma_i+j\pi$};
		\node [style=none] (22) at (8.75, 1.75) {$\sigma' = (-1)^j \sigma$};
		\node [style=Z phase dot] (29) at (1.175, 1) {$\alpha_1$};
		\node [style=Z phase dot] (30) at (1.175, -0.5) {$\alpha_2$};
		\node [style=Z phase dot] (31) at (5.675, -0.5) {$\gamma_2'$};
		\node [style=Z phase dot] (32) at (4.175, 2) {$\beta_1'$};
		\node [style=Z phase dot] (33) at (5.675, 1) {$\gamma_1'$};
		\node [style=Z dot] (36) at (2.425, 2) {};
		\node [style=Z phase dot] (37) at (2.425, 3) {$\sigma'$};
		\node [style=none] (39) at (1.675, 3.5) {};
		\node [style=none] (40) at (3.175, 3.5) {};
		\node [style=none] (41) at (3.175, 1.5) {};
		\node [style=none] (42) at (1.675, 1.5) {};
		\node [style=none] (43) at (6.175, 3) {$\leftarrow$ \textit{phase gadget}};
		\node [style=none] (44) at (-6.325, 2) {};
		\node [style=none] (45) at (-6.325, 0.5) {};
		\node [style=none] (46) at (-2.575, 2) {};
		\node [style=none] (47) at (-2.575, 0.5) {};
		\node [style=none] (48) at (-7.825, 1) {};
		\node [style=none] (49) at (-7.825, -0.5) {};
		\node [style=none] (50) at (-1.325, 1) {};
		\node [style=none] (51) at (-1.325, -0.5) {};
		\node [style=none] (52) at (0.425, 1) {};
		\node [style=none] (53) at (0.425, -0.5) {};
		\node [style=none] (54) at (6.425, 1) {};
		\node [style=none] (55) at (6.425, -0.5) {};
	\end{pgfonlayer}
	\begin{pgfonlayer}{edgelayer}
		\draw [style=hadamard edge] (3) to (4);
		\draw [style=hadamard edge] (3) to (6);
		\draw [style=hadamard edge] (3) to (5);
		\draw [style=hadamard edge] (2) to (5);
		\draw [style=hadamard edge] (2) to (3);
		\draw [style=hadamard edge] (2) to (0);
		\draw [style=hadamard edge] (2) to (1);
		\draw [style=hadamard edge] (6) to (5);
		\draw [style=hadamard edge] (0) to (5);
		\draw [style=hadamard edge] (29) to (33);
		\draw [style=hadamard edge] (33) to (32);
		\draw [style=hadamard edge] (32) to (29);
		\draw [style=hadamard edge] (29) to (31);
		\draw [style=hadamard edge] (31) to (30);
		\draw [style=hadamard edge] (30) to (33);
		\draw [style=hadamard edge] (37) to (36);
		\draw [style=hadamard edge] (36) to (29);
		\draw [style=hadamard edge] (36) to (30);
		\draw [style=hadamard edge] (36) to (32);
		\draw [style=box edge] (39.center) to (40.center);
		\draw [style=box edge] (40.center) to (41.center);
		\draw [style=box edge] (41.center) to (42.center);
		\draw [style=box edge] (42.center) to (39.center);
		\draw [style=gray edge] (44.center) to (46.center);
		\draw [style=gray edge] (46.center) to (47.center);
		\draw [style=gray edge] (47.center) to (45.center);
		\draw [style=gray edge] (45.center) to (44.center);
		\draw (1) to (48.center);
		\draw (0) to (49.center);
		\draw (4) to (50.center);
		\draw (51.center) to (6);
		\draw (53.center) to (30);
		\draw (29) to (52.center);
		\draw (33) to (54.center);
		\draw (55.center) to (31);
	\end{pgfonlayer}
\end{tikzpicture}

%% file: tikz/mbqc_xy.tikz
\begin{tikzpicture}
	\begin{pgfonlayer}{nodelayer}
		\node [style=Z phase dot] (0) at (0, 0) {$\alpha$};
		\node [style=none] (1) at (1.5, 0) {};
	\end{pgfonlayer}
	\begin{pgfonlayer}{edgelayer}
		\draw (0) to (1.center);
	\end{pgfonlayer}
\end{tikzpicture}

%% file: tikz/mbqc_xz.tikz
\begin{tikzpicture}
	\begin{pgfonlayer}{nodelayer}
		\node [style=X phase dot] (0) at (0, 0) {$\alpha$};
		\node [style=none] (1) at (2.25, 0) {};
		\node [style=Z phase dot] (2) at (1.25, 0) {$\frac{\pi}{2}$};
	\end{pgfonlayer}
	\begin{pgfonlayer}{edgelayer}
		\draw (0) to (2);
		\draw (2) to (1.center);
	\end{pgfonlayer}
\end{tikzpicture}

%% file: tikz/mbqc_xz-graphlike.tikz
\begin{tikzpicture}
	\begin{pgfonlayer}{nodelayer}
		\node [style=Z phase dot] (0) at (0, 0) {$\alpha$};
		\node [style=none] (1) at (2.25, 0) {};
		\node [style=Z phase dot] (2) at (1.25, 0) {$\frac{\pi}{2}$};
	\end{pgfonlayer}
	\begin{pgfonlayer}{edgelayer}
		\draw [style=hadamard edge] (0) to (2);
		\draw (2) to (1.center);
	\end{pgfonlayer}
\end{tikzpicture}

%% file: tikz/mbqc_yz.tikz
\begin{tikzpicture}
	\begin{pgfonlayer}{nodelayer}
		\node [style=X phase dot] (0) at (0, 0) {$\alpha$};
		\node [style=none] (1) at (1.25, 0) {};
	\end{pgfonlayer}
	\begin{pgfonlayer}{edgelayer}
		\draw (0) to (1.center);
	\end{pgfonlayer}
\end{tikzpicture}

%% file: tikz/mbqc_yz-graphlike.tikz
\begin{tikzpicture}
	\begin{pgfonlayer}{nodelayer}
		\node [style=Z phase dot] (0) at (0, 0) {$\alpha$};
		\node [style=none] (1) at (1.25, 0) {};
	\end{pgfonlayer}
	\begin{pgfonlayer}{edgelayer}
		\draw [style=hadamard edge] (0) to (1.center);
	\end{pgfonlayer}
\end{tikzpicture}

%% file: content/heuristixz.tex
\newcommand{\wires}{\texttt{\#wires}}

\section{Enhancing reduction of 2-qubit gates}\label{sec:heuristikz}
As seen in \autoref{eq:extracted_circuit_unoptimized} and~\ref{eq:extracted_circuit_optimized}, a Hadamard wire gets extracted to H, CNOT or CZ gates.
The number of Hadamard wires in a graph-like ZX-diagram therefore correlates with the number of 2-qubit gates in the extracted circuit.
The diagram simplification algorithms shown in \autoref{sec:related} focus on eliminating spiders while neglecting -- or even increasing -- the number of Hadamard wires.
Hence, this section introduces methods which additionally minimize the amount of Hadamard wires (ref. as \wires{} in the following).

\begin{wrapfigure}{r}{.44\textwidth}
\vspace{-2.75em}
\begin{equation}\label{eq:bad_cnot_reduction}
	\input{bad_cnot_reduction.tikz}
\end{equation}
\vspace{-2.5em}
\end{wrapfigure}
To do so, it is crucial to examine \textit{where} and \textit{when} local complementation and pivoting are applied. 
Both rules can either increase or decrease \wires{}, depending on the connectivity of the relevant neighbors. 
\autoref{eq:lcompRule} and~\ref{eq:pivotRule} show examples in which \wires{} decreases.
However, as \autoref{eq:pivotGadgetExample} shows it can also increase \wires{} and we easily construct extreme cases like the one shown in \autoref{eq:bad_cnot_reduction}.
Applying local complementation to the central spider with phase $\pi/2$ yields a diagram containing one spider less but a significantly higher \wires. 
Extracting the left diagram with the current version of the PyZX-library produces a circuit with six 2-qubit gates, while the diagram on the right gets extracted as a circuit with $ 21 $ 2-qubit gates.
Generally, applying local complementation on a spider with $ n $ unconnected neighbors leads to $ \frac{n(n+1)}{2} -n $ new wires.
As pivoting involves local complementation on two spiders, the effect usually even worsens for this rule.

To prevent such cases and to guide the simplification process towards a minimal \wires{}, we introduce cost functions for local complementation and pivoting allowing us to calculate \wires{} after rule applications.
We take those as a heuristic for estimating how rule applications change the number of 2-qubit gates and implement decision strategies for diagram simpliciation based on the heuristics.

\subsection{Pivoting and local complementation on spiders with arbitrary phases}\label{sec:heuristikz-arbitrary}
\begin{wrapfigure}{r}{.4\textwidth}
	\vspace{-2.25em}
	\begin{equation}\label{eq:lcomp_on_gadget2}
		\input{lcomp_on_gadget3.tikz}
	\end{equation}
	\vspace{-1.75em}
\end{wrapfigure}

In contrast to the Clifford simplification algorithm from \autoref{sec:related}, we can apply local complementation and pivoting on spiders with \textit{arbitrary} phases. 
Similar to the \textit{Pivot Phase Gadget}~(p2) rule, we can change a spider with non-Clifford phase by a combination of the rules ($ f,i2,i1 $) as in \autoref{eq:lcomp_on_gadget2}.
With that we can apply local complementation on spiders with phase different from $\pm\pi/2$ (this introduces one $ XZ $-spider) and pivoting on pairs of spiders where one/no spider has a phase of $0$ or $\pi$ (this introduces one/two $ YZ $-spiders). Note that this does not change the gflow property (c.f. \cite[Lemma 3.1]{backens2021there}).

\subsection{Local Complementation Heuristic (LCH)}
The costs for local complementation are calculated on the following proposition:

\begin{proposition}
	\noindent
	Let $G = \left(V,E\right)$ be an open graph; $u \in V$ an arbitrary vertex with neighbors \mbox{$N\left(u\right) \subset V$}; $n = \left|N\left(u\right)\right|$ the number of neighbors; and $m$ the number of edges between the neighbors, i.e., 
	\[m = \left|\left\{\left(a,b\right) \in E| a,b \in N\left(u\right)\right\}\right|.\]
	For $G\star u$, $n$ remains the same, but $m$ changes to \mbox{$m' = \triangle_{n-1} - m$}, where $\triangle_{n-1} = \frac{n\left(n-1\right)}{2}$.
\end{proposition}
Hence, the difference in the number of wires after application of the local complementation rule is:
\begin{equation}
	 \left(n+m\right) - \left(n + \left(\triangle_{n-1} - m\right)\right) = 2m - \triangle_{n-1}
\end{equation}
With respect to the phase $ \varphi(u) $ of the spider $ u $, the graph changes as follows:
\begin{itemize}
	\item If $\varphi\left(u\right)=\pm\pi/2$: Remove $ u $ from the graph and eliminate all wires between $ u $ and $ N(u) $.
	\item If $\varphi\left(u\right)$ is non-Clifford:  All wires between $u$ and $N\left(u\right)$ remain and we get an additional wire for the \textit{phase gadget}.
  	\item If $\varphi\left(u\right)$ is $0$ or $\pi$: No phase gadget is needed and we can use the $ \pi $-copy rule.
\end{itemize}
The $ LCH $ is calculated as follows:
\begin{equation}
	LCH\left(u\right) = \begin{cases}
		2m - \triangle_{n-1} + n & \text{if } \varphi\left(u\right) = \pm\frac{\pi}{2} \\
		2m - \triangle_{n-1} & \text{if } \varphi\left(u\right) = k\cdot\pi, k\in \mathbb{Z} \\
		2m - \triangle_{n-1} - 1 & \text{otherwise }
	\end{cases}
\end{equation}

\subsection{Pivoting Heuristic (PH)}
We calculate the upper bound of new connections with the sets $ A,B,C $ (see \autoref{sec:background-graph}): 
\begin{equation}C_{max} = |A|\cdot|B|+|A|\cdot|C|+|B|\cdot|C|\end{equation}
We denote the number of neighbors of $u$ and $v$ by $n_u = |N(u)|$ and $n_v = |N(v)|$, respectively,
and the number of edges between neighbors of different sets as $ m $.

The changes of the graph $ G $ to $ G\land uv $ have the following cases ($ j,k\in \mathbb{Z} $):
\begin{enumerate}[label=(C\arabic*)]
	\item $ \varphi\left(u\right) = j\cdot\pi, \varphi\left(v\right) = k\cdot\pi $: 
		If both spiders have a phase of $0$ or $\pi$, all connections between $\left\{u,v\right\}$ and $N\left(u\right) \cup N\left(v\right)$ are eliminated.
	\item $ \varphi\left(u\right) = j\cdot\pi, \varphi\left(v\right) \neq k\cdot\pi $: 
		If $v$ becomes a phase gadget and $u$ gets eliminated, all neighbors of $u$ get connected to $v$ and we have an additional wire for the phase gadget.
	\item $ \varphi\left(u\right) \neq j\cdot\pi, \varphi\left(v\right) = k\cdot\pi $: 
        If $u$ becomes a phase gadget and $v$ gets eliminated, all neighbors of $v$ get connected to $u$ and we have an additional wire for the phase gadget.
	\item $ \varphi\left(u\right) \neq j\cdot\pi, \varphi\left(v\right) \neq k\cdot\pi $: 
		If both spiders become phase gadgets, all neighbors of $u$ get connected to $v$ and all neighbors of $v$ get connected to $u$. Furthermore, $u$ gets connected to $v$ again and we have two more wires for the phase gadgets.
\end{enumerate}
With these conditions, the $ PH $ is calculated as follows:
\begin{equation}\small
	PH\left(u,v\right) = \begin{cases}
		2m - C_{max}  + n_u + n_v - 1 & \text{for (C1)}\\
		2m - C_{max}  + n_v - 1 & \text{for (C2)}\\
		2m - C_{max}  + n_u - 1 & \text{for (C3)}\\
		2m - C_{max} - 2 & \text{for (C4)} \\ 
	\end{cases}
\end{equation}

\subsection{Decision strategies}\label{sec:decision_strategies}
With the two heuristics ($ LCH, PH $) at hand we can now implement different strategies to decide where and when local complementation or pivoting are applied during the simplification.
A single simplification step in our procedure consists of the following actions:
\begin{enumerate}
	\item Filter all possible rule applications of the current ZX-diagram.
	\item Select rule according to selection strategy (see below).
	\item Apply rule on the ZX-diagram.
\end{enumerate}
For filtering rule applications we can specify a lower bound for the heuristic, e.g., $ LCH\text{ or }PH =-5$ says that we do not consider rule applications which increase \wires{} by more than five. 
We can also specify whether rule applications are allowed on boundary spiders (c.f. \autoref{app:further-rules}) and whether rules are allowed on arbitrary phased spiders. 
For selecting a rule we implemented two different strategies:
\begin{itemize}
	\item \textbf{Random selection}: Rules are chosen by a random coin flip.
	\item \textbf{Greedy selection}: Chooses the rule application which maximally decreases \wires.
\end{itemize}
Each algorithm terminates if we allow only rule applications with a $ LCH/PH >0$ and when there is no rule left that decreases \wires. 
They also terminate if we allow negative gains ($ LCH/PH \leq 0 $) and restrict the matches to interior spiders that do not generate new spiders. 
This is the case in standard local complementation on a spider with phase $\pm\pi/2$ and pivoting on a pair of spiders with phase $0$ or $\pi$. 
The algorithm eliminates at least one spider in every step and terminates when neither interior spiders with phase $\pm\pi/2$ nor pairs with phase $0$ or $\pi$ are left.


On the other hand, allowing rule applications on spiders of arbitrary phases which increase \wires{} may result in loops and therefore no termination.
For such cases we only allow rule applications which increase \wires{} on spiders present since the very beginning of our simplification procedure. On newly generated spiders only rules which decrease \wires{} are allowed.

%% file: tikz/bad_cnot_reduction.tikz
\begin{tikzpicture}
	\begin{pgfonlayer}{nodelayer}
		\node [style=none] (0) at (-6.5, 1.25) {};
		\node [style=none] (1) at (-6.5, 2.5) {};
		\node [style=none] (2) at (-6.5, 0) {};
		\node [style=none] (3) at (-6.5, -1.25) {};
		\node [style=Z phase dot] (4) at (-5.75, 2.5) {$\pi$};
		\node [style=Z phase dot] (9) at (-4, 0) {$\frac{\pi}{2}$};
		\node [style=Z phase dot] (13) at (-5.75, -1.25) {$\text{-}\frac{\pi}{2}$};
		\node [style=Z phase dot] (21) at (-4, 1.25) {$\frac{\pi}{4}$};
		\node [style=Z phase dot] (24) at (-4, 2.5) {$\frac{\pi}{4}$};
		\node [style=Z dot] (30) at (-2.5, 1.25) {};
		\node [style=Z dot] (32) at (-2.5, 2.5) {};
		\node [style=none] (33) at (-1.75, 2.5) {};
		\node [style=none] (34) at (-1.75, 1.25) {};
		\node [style=none] (35) at (-1.75, 0) {};
		\node [style=none] (36) at (-1.75, -1.25) {};
		\node [style=Z phase dot] (39) at (-2.5, -1.25) {$\pi$};
		\node [style=Z dot] (57) at (-2.5, 0) {};
		\node [style=Z phase dot] (58) at (-5.75, 1.25) {$\text{-}\frac{\pi}{2}$};
		\node [style=Z phase dot] (59) at (-4, -1.25) {$\frac{\pi}{4}$};
		\node [style=Z dot] (60) at (-5.75, 0) {};
		\node [style=none] (61) at (-0.75, 0.75) {$\overset{(lc)}{=}$};
		\node [style=none] (63) at (0.25, 1.25) {};
		\node [style=none] (64) at (0.25, 2.5) {};
		\node [style=none] (65) at (0.25, 0) {};
		\node [style=none] (66) at (0.25, -1.25) {};
		\node [style=Z phase dot] (67) at (1.25, 2.5) {$\frac{\pi}{2}$};
		\node [style=Z phase dot] (69) at (1.25, -1.25) {$\pi$};
		\node [style=Z phase dot] (70) at (2.75, 1.25) {$\frac{\pi}{4}$};
		\node [style=Z phase dot] (71) at (2.75, 2.5) {$\frac{\pi}{4}$};
		\node [style=Z phase dot] (72) at (4.25, 1.25) {$\text{-}\frac{\pi}{2}$};
		\node [style=Z phase dot] (73) at (4.25, 2.5) {$\text{-}\frac{\pi}{2}$};
		\node [style=none] (74) at (5.25, 2.5) {};
		\node [style=none] (75) at (5.25, 1.25) {};
		\node [style=none] (76) at (5.25, 0) {};
		\node [style=none] (77) at (5.25, -1.25) {};
		\node [style=Z phase dot] (78) at (4.25, -1.25) {$\frac{\pi}{2}$};
		\node [style=Z phase dot] (79) at (4.25, 0) {$\text{-}\frac{\pi}{2}$};
		\node [style=Z phase dot] (80) at (1.25, 1.25) {$\pi$};
		\node [style=Z phase dot] (81) at (2.75, -1.25) {$\frac{\pi}{4}$};
		\node [style=Z phase dot] (82) at (1.25, 0) {$\text{-}\frac{\pi}{2}$};
		\node [style=none] (83) at (-4.5, 0.5) {};
		\node [style=none] (84) at (-4.5, -0.5) {};
		\node [style=none] (85) at (-3.5, 0.5) {};
		\node [style=none] (86) at (-3.5, -0.5) {};
	\end{pgfonlayer}
	\begin{pgfonlayer}{edgelayer}
		\draw (33.center) to (32);
		\draw (34.center) to (30);
		\draw [style=hadamard edge] (32) to (24);
		\draw (4) to (1.center);
		\draw (36.center) to (39);
		\draw (57) to (35.center);
		\draw [style=hadamard edge] (21) to (30);
		\draw [style=hadamard edge] (4) to (24);
		\draw [style=hadamard edge] (9) to (30);
		\draw [style=hadamard edge] (57) to (9);
		\draw (3.center) to (13);
		\draw (58) to (0.center);
		\draw [style=hadamard edge] (58) to (21);
		\draw [style=hadamard edge] (9) to (58);
		\draw [style=hadamard edge] (9) to (32);
		\draw [style=hadamard edge] (9) to (39);
		\draw [style=hadamard edge] (9) to (13);
		\draw [style=hadamard edge] (9) to (4);
		\draw [style=hadamard edge] (13) to (59);
		\draw [style=hadamard edge] (59) to (39);
		\draw [style=hadamard edge] (60) to (9);
		\draw (60) to (2.center);
		\draw (74.center) to (73);
		\draw (75.center) to (72);
		\draw [style=hadamard edge] (73) to (71);
		\draw (67) to (64.center);
		\draw (77.center) to (78);
		\draw (79) to (76.center);
		\draw [style=hadamard edge] (70) to (72);
		\draw [style=hadamard edge] (67) to (71);
		\draw (66.center) to (69);
		\draw (80) to (63.center);
		\draw [style=hadamard edge] (80) to (70);
		\draw [style=hadamard edge] (69) to (81);
		\draw [style=hadamard edge] (81) to (78);
		\draw (82) to (65.center);
		\draw [style=hadamard edge] (82) to (79);
		\draw [style=hadamard edge, bend left] (67) to (73);
		\draw [style=hadamard edge, bend left=45] (80) to (72);
		\draw [style=hadamard edge, bend right] (69) to (78);
		\draw [style=hadamard edge] (69) to (79);
		\draw [style=hadamard edge] (69) to (72);
		\draw [style=hadamard edge] (69) to (73);
		\draw [style=hadamard edge] (69) to (82);
		\draw [style=hadamard edge, bend left] (69) to (80);
		\draw [style=hadamard edge, bend left=45] (69) to (67);
		\draw [style=hadamard edge] (67) to (80);
		\draw [style=hadamard edge, bend right] (67) to (82);
		\draw [style=hadamard edge] (80) to (82);
		\draw [style=hadamard edge] (73) to (72);
		\draw [style=hadamard edge] (72) to (79);
		\draw [style=hadamard edge] (79) to (78);
		\draw [style=hadamard edge, bend left] (73) to (79);
		\draw [style=hadamard edge, bend left] (72) to (78);
		\draw [style=hadamard edge, bend left=45] (73) to (78);
		\draw [style=hadamard edge] (78) to (67);
		\draw [style=hadamard edge] (78) to (80);
		\draw [style=hadamard edge] (78) to (82);
		\draw [style=hadamard edge] (82) to (72);
		\draw [style=hadamard edge] (82) to (73);
		\draw [style=hadamard edge] (79) to (80);
		\draw [style=hadamard edge] (79) to (67);
		\draw [style=hadamard edge] (67) to (72);
		\draw [style=hadamard edge] (80) to (73);
		\draw [style=gray edge] (83.center) to (85.center);
		\draw [style=gray edge] (85.center) to (86.center);
		\draw [style=gray edge] (86.center) to (84.center);
		\draw [style=gray edge] (84.center) to (83.center);
	\end{pgfonlayer}
\end{tikzpicture}

%% file: tikz/lcomp_on_gadget3.tikz
\begin{tikzpicture}
	\begin{pgfonlayer}{nodelayer}
		\node [style=Z phase dot] (0) at (-2, -1.75) {$\alpha_3$};
		\node [style=Z phase dot] (1) at (-2, 0) {$\pm\frac{\pi}{2}$};
		\node [style=Z phase dot] (2) at (-0.5, -0.25) {$\alpha_1$};
		\node [style=Z phase dot] (3) at (-0.5, -1.75) {$\alpha_2$};
		\node [style=none] (18) at (0.25, -1.1) {$\overset{(lc)}{=}$};
		\node [style=Z phase dot] (19) at (3.5, -0.25) {$\alpha_1\mp\frac{\pi}{2}$};
		\node [style=Z phase dot] (20) at (3.5, -1.75) {$\alpha_2\mp\frac{\pi}{2}$};
		\node [style=Z phase dot] (21) at (1.5, -1.75) {$\alpha_{3}\mp\frac{\pi}{2}$};
		\node [style=Z phase dot] (36) at (-2, 1.5) {$\beta\mp\frac{\pi}{2}$};
		\node [style=Z phase dot] (37) at (1.5, -0.25) {$\mp\frac{\pi}{2}$};
		\node [style=Z phase dot] (38) at (-4.25, 0) {$\alpha_1$};
		\node [style=Z phase dot] (39) at (-5.75, 0) {$\beta$};
		\node [style=Z phase dot] (40) at (-5.75, -1.75) {$\alpha_2$};
		\node [style=Z phase dot] (41) at (-4.25, -1.75) {$\alpha_3$};
		\node [style=none] (44) at (-3.25, -1.1) {$\overset{(f,i2,i1)}{=}$};
		\node [style=Z dot] (45) at (-2, 0.75) {};
		\node [style=Z phase dot] (46) at (1.5, 1) {$\beta\mp\frac{\pi}{2}$};
		\node [style=none] (47) at (-6.25, 0.5) {};
		\node [style=none] (48) at (-6.25, -0.5) {};
		\node [style=none] (49) at (-3.75, 0.5) {};
		\node [style=none] (50) at (-3.75, -0.5) {};
		\node [style=none] (51) at (-2.525, 0.5) {};
		\node [style=none] (52) at (-2.525, -0.5) {};
		\node [style=none] (53) at (-1.475, 0.5) {};
		\node [style=none] (54) at (-1.475, -0.5) {};
	\end{pgfonlayer}
	\begin{pgfonlayer}{edgelayer}
		\draw [style=hadamard edge] (1) to (0);
		\draw [style=hadamard edge, in=15, out=-165] (1) to (2);
		\draw [style=hadamard edge] (1) to (3);
		\draw [style=hadamard edge] (21) to (19);
		\draw [style=hadamard edge] (20) to (19);
		\draw [style=hadamard edge] (37) to (19);
		\draw [style=hadamard edge] (37) to (20);
		\draw [style=hadamard edge] (37) to (21);
		\draw [style=hadamard edge] (39) to (38);
		\draw [style=hadamard edge] (39) to (40);
		\draw [style=hadamard edge] (39) to (41);
		\draw [style=hadamard edge] (36) to (45);
		\draw [style=hadamard edge] (45) to (1);
		\draw [style=hadamard edge] (46) to (37);
		\draw [style=hadamard edge] (40) to (41);
		\draw [style=hadamard edge] (0) to (3);
		\draw [style=gray edge] (47.center) to (49.center);
		\draw [style=gray edge, in=90, out=-90] (49.center) to (50.center);
		\draw [style=gray edge, in=360, out=180] (50.center) to (48.center);
		\draw [style=gray edge, in=270, out=90, looseness=1.50] (48.center) to (47.center);
		\draw [style=gray edge] (51.center) to (53.center);
		\draw [style=gray edge, in=90, out=-90] (53.center) to (54.center);
		\draw [style=gray edge, in=360, out=180] (54.center) to (52.center);
		\draw [style=gray edge, in=270, out=90, looseness=1.50] (52.center) to (51.center);
	\end{pgfonlayer}
\end{tikzpicture}

%% file: content/neighbor-unfusion.tex
\section{New optimization rule: Neighbor Unfusion}

As shown in \autoref{eq:pivotGadgetExample}, the application of local complementation ($ lc $) and pivoting ($ p $) on spiders with many neighbors can not only decrease but also increase \wires.
The heuristics shown in \autoref{sec:heuristikz} may help to identify and prevent extreme cases as in \autoref{eq:bad_cnot_reduction}.
However, spiders that are measured in $YZ$- or $XZ$-plane (c.f. \autoref{sec:related-extraction}) require special attention:
When extracting a spider in $ YZ $-plane (e.g., the empty spider of the phase gadget in \autoref{eq:pivot_phase_gadget} and~\ref{eq:pivotGadgetExample}), pivoting has to be applied to maintain the focused gflow property.
The same happens for spiders in $ XZ $-plane, but they are resolved by local complementation~\cite{backens2021there}.
This affects \wires{} \textit{after} the simplification and a simplified diagram containing some spiders in $YZ$- and $XZ$-plane may result in an expensive circuit.

However, when applying either rule on diagrams with arbitrary spiders as discussed in \autoref{sec:heuristikz-arbitrary}), spiders in  $YZ$- or $XZ$-plane are generated.
We introduce the \textit{neighbor unfusion} ($ nu $) rule, which allows $ lc $ and $ p $ on such arbitrary-phase spiders without introducing spiders in  $YZ$- or $XZ$-plane.

\begin{wrapfigure}{r}{.42\textwidth}
	\vspace{-1.875em}
	\begin{equation}\label{eq:neighbor_unfusion_basic}
		\input{neighbor_unfusion_basic.tikz}
	\end{equation}
\vspace{-1.75em}
\end{wrapfigure}
\textit{Neighbor unfusion} combines the fusion ($ f $) and identity rules ($ i1,i2 $) as shown in \autoref{eq:neighbor_unfusion_basic}.
If a spider with phase $\alpha$ is connected to a neighbor, we change its phase to an arbitrary phase $\gamma$ by inserting an empty spider and a spider with phase $\alpha-\gamma$ between the spider and its neighbor.
It allows changing the phase of an arbitrary spider to $\gamma=\pm\frac{\pi}{2}$ and thus local complementation and removal of spiders with arbitrary phases.

\begin{wrapfigure}{r}{.465\textwidth}
	\vspace{-3.375em}
	\begin{equation}\label{eq:lcompRuleNeighbors2}
		\input{lcompRuleNeighbors3.tikz}
	\end{equation}
\vspace{-1.75em}
\end{wrapfigure}
For illustration, we apply neighbor unfusion ($ nu $) to the example of \autoref{eq:lcomp_on_gadget2}.
We can move the $ \beta $ spider to any direction (in \autoref{eq:lcompRuleNeighbors2} towards $ \alpha_1 $), so it is then not affected by the application of local complementation~($ lc $).
Comparing  \autoref{eq:lcomp_on_gadget2} and \autoref{eq:lcompRuleNeighbors2} we see that we not only reduce \wires{}, but also prevent the generation of a spider in $ XZ $-plane.
However, the neighbor unfusion rule sometimes leads to diagrams which do not have focused gflow property. This is due to the insertion of the empty spider and the spider with phase $\alpha-\gamma$ in \ref{eq:neighbor_unfusion_basic}. 
We observed that this problem does not occur if the spiders with phase $\alpha$ and $\beta$ get extracted to the same qubit.
Currently, we find such pairs of spiders by using the flow hierarchy of the \textit{maximally delayed gflow} of the diagram (c.f. \cite{backens2021there}), which is quite costly, because we need to recalculate the gflow at each simplification step. Therefore,  diagram simplification with neighbor unfusion has a much higher runtime than the other simplification procedures.
It is an open question whether neighbour unfusion can only destroy the focused gflow property, or also more general flow properties like gflow or Pauli flow.

%% file: tikz/neighbor_unfusion_basic.tikz
\begin{tikzpicture}
	\begin{pgfonlayer}{nodelayer}
		\node [style=none] (0) at (-6.5, 1) {};
		\node [style=none] (1) at (-4.5, 1) {};
		\node [style=Z phase dot] (2) at (-5.5, 0) {$\alpha$};
		\node [style=none] (3) at (-4.5, -1) {};
		\node [style=none] (4) at (-6.5, -1) {};
		\node [style=none] (5) at (-4.5, -1) {};
		\node [style=none] (6) at (-4, 1) {};
		\node [style=none] (7) at (-2, 1) {};
		\node [style=Z phase dot] (8) at (-3, 0) {$\beta$};
		\node [style=none] (9) at (-2, -1) {};
		\node [style=none] (10) at (-4, -1) {};
		\node [style=none] (11) at (-2, -1) {};
		\node [style=none] (12) at (-5.5, 0.825) {$\ldots$};
		\node [style=none] (13) at (-3, 0.825) {$\ldots$};
		\node [style=none] (14) at (-3, -0.825) {$\ldots$};
		\node [style=none] (15) at (-5.5, -0.825) {$\ldots$};
		\node [style=none] (16) at (-1.5, 0.25) {$\overset{(nu)}{=}$};
		\node [style=none] (17) at (-1, 1) {};
		\node [style=none] (18) at (1, 1) {};
		\node [style=Z phase dot] (19) at (0, 0) {$\gamma$};
		\node [style=none] (20) at (1, -1) {};
		\node [style=none] (21) at (-1, -1) {};
		\node [style=none] (22) at (1, -1) {};
		\node [style=none] (23) at (2.75, 1) {};
		\node [style=none] (24) at (4.75, 1) {};
		\node [style=Z phase dot] (25) at (3.75, 0) {$\beta$};
		\node [style=none] (26) at (4.75, -1) {};
		\node [style=none] (27) at (2.75, -1) {};
		\node [style=none] (28) at (4.75, -1) {};
		\node [style=none] (29) at (0, 0.825) {$\ldots$};
		\node [style=none] (30) at (3.75, 0.825) {$\ldots$};
		\node [style=none] (31) at (3.75, -0.825) {$\ldots$};
		\node [style=none] (32) at (0, -0.825) {$\ldots$};
		\node [style=Z dot] (33) at (1, 0) {};
		\node [style=Z phase dot] (34) at (2.225, 0) {$\alpha-\gamma$};
	\end{pgfonlayer}
	\begin{pgfonlayer}{edgelayer}
		\draw [style=hadamard edge, in=-60, out=90] (5.center) to (2);
		\draw [style=hadamard edge, in=-90, out=60] (2) to (1.center);
		\draw [style=hadamard edge, in=-90, out=135, looseness=1.25] (2) to (0.center);
		\draw [style=hadamard edge, in=90, out=-135] (2) to (4.center);
		\draw [style=hadamard edge, in=-60, out=90] (11.center) to (8);
		\draw [style=hadamard edge, in=-90, out=60] (8) to (7.center);
		\draw [style=hadamard edge, in=-90, out=135, looseness=1.25] (8) to (6.center);
		\draw [style=hadamard edge, in=90, out=-135] (8) to (10.center);
		\draw [style=hadamard edge] (2) to (8);
		\draw [style=hadamard edge, in=-60, out=90] (22.center) to (19);
		\draw [style=hadamard edge, in=-90, out=60] (19) to (18.center);
		\draw [style=hadamard edge, in=-90, out=135, looseness=1.25] (19) to (17.center);
		\draw [style=hadamard edge, in=90, out=-135] (19) to (21.center);
		\draw [style=hadamard edge, in=-60, out=90] (28.center) to (25);
		\draw [style=hadamard edge, in=-90, out=60] (25) to (24.center);
		\draw [style=hadamard edge, in=-90, out=135, looseness=1.25] (25) to (23.center);
		\draw [style=hadamard edge, in=90, out=-135] (25) to (27.center);
		\draw [style=hadamard edge] (19) to (33);
		\draw [style=hadamard edge] (33) to (34);
		\draw [style=hadamard edge] (34) to (25);
	\end{pgfonlayer}
\end{tikzpicture}

%% file: tikz/lcompRuleNeighbors3.tikz
\begin{tikzpicture}
	\begin{pgfonlayer}{nodelayer}
		\node [style=Z phase dot] (86) at (-2.5, -1) {$\alpha_3$};
		\node [style=Z phase dot] (87) at (-2.5, 1) {$\pm\frac{\pi}{2}$};
		\node [style=Z dot] (88) at (-1.6, 1) {};
		\node [style=Z phase dot] (89) at (-0.75, -1) {$\alpha_2$};
		\node [style=none] (90) at (2, 0) {$\overset{(lc)}{=}$};
		\node [style=Z phase dot] (94) at (-0.35, 1) {$\beta\mp\frac{\pi}{2}$};
		\node [style=Z phase dot] (96) at (-4.25, 1) {$\alpha_1$};
		\node [style=Z phase dot] (97) at (-6, 1) {$\beta$};
		\node [style=Z phase dot] (98) at (-6, -1) {$\alpha_2$};
		\node [style=Z phase dot] (99) at (-4.25, -1) {$\alpha_3$};
		\node [style=none] (100) at (-3.5, 0) {$\overset{(nu)}{=}$};
		\node [style=Z phase dot] (103) at (1.15, 1) {$\alpha_1$};
		\node [style=Z phase dot] (104) at (2.75, -1) {$\alpha_3\mp\frac{\pi}{2}$};
		\node [style=Z phase dot] (107) at (4.75, -1) {$\alpha_2\mp\frac{\pi}{2}$};
		\node [style=Z phase dot] (108) at (4.25, 1) {$\beta\mp\frac{\pi}{2}$};
		\node [style=Z phase dot] (109) at (5.75, 1) {$\alpha_1$};
		\node [style=Z phase dot] (111) at (2.75, 1) {$\mp\frac{\pi}{2}$};
		\node [style=none] (112) at (-3.05, 1.5) {};
		\node [style=none] (113) at (-3.05, 0.5) {};
		\node [style=none] (116) at (-1.9, 1.5) {};
		\node [style=none] (117) at (-1.9, 0.5) {};
		\node [style=none] (118) at (-6.5, 1.5) {};
		\node [style=none] (119) at (-6.5, 0.5) {};
		\node [style=none] (120) at (-3.75, 1.5) {};
		\node [style=none] (121) at (-3.75, 0.5) {};
	\end{pgfonlayer}
	\begin{pgfonlayer}{edgelayer}
		\draw [style=hadamard edge] (87) to (86);
		\draw [style=hadamard edge] (87) to (88);
		\draw [style=hadamard edge] (87) to (89);
		\draw [style=hadamard edge] (97) to (96);
		\draw [style=hadamard edge] (97) to (98);
		\draw [style=hadamard edge] (97) to (99);
		\draw [style=hadamard edge] (98) to (99);
		\draw [style=hadamard edge] (86) to (89);
		\draw [style=hadamard edge] (94) to (88);
		\draw [style=hadamard edge] (94) to (103);
		\draw [style=hadamard edge] (108) to (109);
		\draw [style=hadamard edge] (104) to (111);
		\draw [style=hadamard edge] (111) to (107);
		\draw [style=hadamard edge] (111) to (108);
		\draw [style=gray edge] (112.center) to (116.center);
		\draw [style=gray edge] (116.center) to (117.center);
		\draw [style=gray edge] (117.center) to (113.center);
		\draw [style=gray edge] (113.center) to (112.center);
		\draw [style=gray edge] (118.center) to (120.center);
		\draw [style=gray edge, in=90, out=-90] (120.center) to (121.center);
		\draw [style=gray edge, in=360, out=180] (121.center) to (119.center);
		\draw [style=gray edge, in=270, out=90, looseness=1.50] (119.center) to (118.center);
	\end{pgfonlayer}
\end{tikzpicture}

%% file: content/evaluation.tex

\newcommand{\cellh}[1]{\cellcolor[cmyk]{0.25,0,0.5,0}\textbf{#1}}
\newcommand{\cellb}[1]{\cellcolor[cmyk]{0, 0.5, 0.5, 0}#1}
\newcommand{\tket}{t$ \ket{\text{ket}} $}

\begin{table}[t]
		\caption{Circuit metrics for original benchmark circuits, Post-optimization metrics of the standard Clifford simplification~\cite{duncan2020graph},  PyZX \cite{kissinger2020reducing},  Nam et al.\cite{nam2018automated}, our heuristic-based simplification method, and the combined approach of Nam et al. and our heuristic-based methods.
			Only our best optimization is presented: 1) Greedy, 2) Random, 3) Greedy with neighbor unfusion, 4) Random with neighbor unfusion, the lower bound for the heuristics is denoted in brackets.
			If the best PyZX result is achieved by the  \tket-library, the respective cell is marked with $ \star $. 
			The best results of each metric in each row are marked.}
		\label{tab:clifford_simp_compare}\label{tab:best_results}
		\footnotesize
		\input{tabulars/tpar_all_simp}
\end{table}

\section{Evaluation}\label{sec:evaluation}
We evaluate our heuristic-based simplification algorithms on a set of circuits first used in \cite{amy2014polynomial}. 
They implement various arithmetic problems as quantum circuits and were used as a benchmark set for comparing different optimization strategies~\cite{nam2018automated,kissinger2020reducing}.
We use it to compare our heuristic-based approaches with the Clifford simplification algorithm described in~\cite{duncan2020graph}, and some of the best results reported for circuit optimizations with~\cite{kissinger2020reducing} and without using ZX-calculus~\cite{nam2018automated}. 
We also investigate how ZX-calculus based approaches perform when using the TODD-algorithm~\cite{heyfron2018efficient} for additional T gate reduction. 
\subsection{Implementation}\label{sec:implementation}
With the exception of~\cite{kissinger2020reducing} -- which omits simplifying the diagram (step 3) and extraction (step 4) -- we use the following pipeline for ZX-calculus based optimization algorithms:
\begin{enumerate}
	\item Optimize circuit using gate cancellation and commutation.
	\item Transform circuit to ZX-diagram and apply phase teleportation to reduce T-count (as in~\cite{kissinger2020reducing}).
	\item Simplify ZX-diagram (standard Clifford or heuristic-based simplification).
	\item Extract circuit from ZX-diagram.
	\item Optimize circuit as in step 1.
\end{enumerate}
Since our heuristic-based algorithms do not reduce T gates, we always apply the phase teleportation in step 2 since this reduces T gates as far as currently possible with ZX-calculus. 
This ensures comparable results regarding the 2-qubit gate count against non-ZX-calculus based approaches that optimize all types of gates.
We implemented our algorithms in a clone of the PyZX library which also contains our optimized circuits in the OpenQASM format\footnote{\url{https://github.com/mnm-team/pyzx-heuristics}}. All results were proven to be correct by checking whether the optimized circuit together with the adjoint of the original circuit can be reduced to the identity. 

\begin{wraptable}[14]{R}{.5\textwidth}
	\vspace{-1.25em}
	\caption{Circuit metrics for the original benchmark circuits, Post-optimization metrics for PyZX+TODD and of our heuristic-based algorithms+TODD.}
	\label{tab:todd_compare}
	\vspace{-1em}
	\input{tabulars/tpar_todd}
\end{wraptable}

\subsection{Results}
For each circuit we compare the total gate count $\Sigma$ and the 2-qubit gate count $2Q$. The results are summarized in \autoref{tab:clifford_simp_compare} and~\ref{tab:todd_compare}: 
Their columns show circuit name, metrics of the original circuit of the benchmark, metrics of one (or more) existing optimization algorithms and the metrics of the best performing heuristic-based algorithm. 
For the latter, we denote the simplification strategy achieving the best result in the last column:
\begin{enumerate*}
	\item Greedy,
	\item Random,
	\item Greedy with neighbor unfusion,
	\item Random with neighbor unfusion.
\end{enumerate*}

As a very first result, the last column in the ``Heuristic Algorithm'' section of \autoref{tab:best_results} prominently indicates the great value of neighbor unfusion (Alg.~3 and~4), as it achieves the best performance of our heuristics in $ >70\% $ of the cases.

We now compare our heuristic-based simplifications following against other ZX-calculus based optimizations in \autoref{tab:best_results}.
For most circuits our heuristic-based simplification clearly outperforms the standard Clifford simplification~\cite{duncan2020graph}, both in total and 2-qubit gate reduction. 
Moreover, while our approaches almost always decrease circuit metrics, the standard approach often yields circuits with higher metrics than the original circuit (e.g.,``CSLA-MUX$ _3 $'', ``GF($ 2^6 $)-Mult'').
Especially for 2-qubit gates our approaches \textit{decrease} 2-qubit gate count by $16\%$, while the standard approach even \textit{increases} the count by $22\%$.
In a direct comparison our approaches have up to 33\% (``Toff-NC${}_4$'') fewer total and 47\% (``Mod-Mult${}_{55}$'') fewer 2-qubit gates than the standard Clifford approach.

Second, we compare against the best available PyZX implementation~\cite{kissinger2020reducing} and the recommended optimization pipeline of the \tket-library~\cite{sivarajah2020t} with the routines \texttt{PauliSimp} and \texttt{FullPeepholeOptimize}, which use similar strategies. 
The column ``PyZX/\tket'' in \autoref{tab:clifford_simp_compare} shows the best optimization results for both implementations and $\star$ indicates results from \tket.
Except for two circuits (``Mod $5_4$'' and ``Toff-Barenco${}_3$''), our algorithms outperform all ZX-calculus based algorithms in terms of total gate count and 2-qubit gate count.

Third, \autoref{tab:best_results} also shows our result in comparison to the cutting-edge non-ZX-calculus based algorithm from Nam et al.~\cite{nam2018automated}. 
It can be seen that the algorithm from \cite{nam2018automated} outperforms any ZX-calculus based algorithm for most circuits.
Still, we were able to achieve better results for the circuits ``VBE-Adder${}_3$'' and ``Mod-Mult${}_{55}$''. 
Note that we did not compare for the ``QCLA-Mod${}_7$'' circuit, because~\cite{hietala2021verified} reports that the optimized circuit from ~\cite{nam2018automated} does not correspond to the original.

Last, the rightmost columns of \autoref{tab:best_results} show a combination of Nam et al's approach with ours. 
We use the output circuits from the Nam et al. optimization as input for our algorithms and observe that we achieve equally good or better results for almost all circuits. 
The larger the circuit, the more significantly this combination improves the previous best known results.
Most notably, we improve the total count of the ``Adder${}_8$'' circuit by more than 15\% 
and the 2-qubit gate count by more than 12\%.


Apart from the 2-qubit gate count, the T gate count of a quantum circuit is an important metric, since T-gates are more complex to implement for an error-corrected quantum computer.
Therefore, we compare our algorithms to the other ZX-calculus based approaches using the TODD algorithm as optimization step 1). 
It is designed to reduce T gate count by introducing ancilla qubits, but sometimes also reduces T-gates in the ancilla-free case. 
\autoref{tab:todd_compare} shows those benchmarks circuits where the combination of TODD and a ZX-calculus based algorithm reduces T gate count even more compared to the best result in \autoref{tab:best_results}. 
We compare the best combination of our heuristic-based algorithm and TODD against the best combination of an existing ZX-calculus based algorithm and TODD.

While we observe a general increase in 2-qubit and total gates using TODD algorithm, our best algorithm yields better results than the existing ZX-calculus based algorithms in every case.

%% file: tabulars/tpar_all_simp.tex
	\setlength{\tabcolsep}{0.6mm}
	\centering
	\begin{tabular}{l|c|c|c|c|c|c|c|c|c|c|c|c|c|c|c}
		\toprule
		\rule{0pt}{1\normalbaselineskip} \multirow{2}{1pt}{\textbf{Circuit}} & \multicolumn{2}{c|}{\textbf{Original}} & \multicolumn{2}{c|}{\textbf{Clifford algorithm}} & \multicolumn{2}{c|}{\textbf{PyZX/\tket${}^\star$}} & \multicolumn{2}{c|}{\textbf{Nam et al.}} & \multicolumn{3}{c|}{\textbf{Heuristic algorithm}} & \multicolumn{3}{c|}{\textbf{Nam+Heuristic}} &  \\
		                                                                     & $\Sigma$ &            $2Q$             & $\Sigma$  &                 $2Q$                 &      $\Sigma$      &             $2Q$              &      $\Sigma$      &        $2Q$         &  $\Sigma$   &    $2Q$     &         Alg.          &   $\Sigma$   &     $2Q$     &     Alg.      &  \\ \midrule
Mod $5_4$              &    63    &             28     &    36     &   21  & \cellh{24}$^\star$ & \cellh{12}$^\star$&         51         &         28          &     41      &     23      &     2(-20)  & 38 & 23 &       3(1)       &  \\
VBE-Adder${}_3$        &   150    &             70     &    116    &   59  &        101         &         54        &         89         &         50          & \cellh{87}  & \cellh{42}  &     3(1)  & \cellh{87} & \cellh{42} &       4(1)       &  \\
CSLA-MUX${}_3$         &   170    &             80     &    177    &   97  &        156         &         75        &    \cellh{155}     &         70          & \cellh{155} &     74      &     3(-5)  &     156      &  \cellh{67}  &  3(1)       &  \\
CSUM-MUX${}_3$         &   420    &             168    &    455    &  271  &    327$^\star$     &     158$^\star$   &    \cellh{266}     &     \cellh{140}     &     303     &     150     &     3(1)  & \cellh{266}  & \cellh{140}  &  1(1)       &  \\
QCLA-Com${}_7$         &   443    &             186    &    397    &  223  &        316         &         148       &        284         &     \cellh{132}     &     295     &     138     &     4(-5)  & \cellh{275}  & \cellh{132}  &  1(1)       &  \\
QCLA-Mod${}_7$         &   884    &             382    &    903    &  475  &        717         &         324       &         -          &          -          & \cellh{705} & \cellh{311} &     4(-20)  &      -       &      -       &  -       &  \\
QCLA-Adder${}_{10}$    &   521    &             233    &    562    &  305  &        435         &         199       &        399         &         183         &     417     &     193     &     4(-20)  & \cellh{398}  & \cellh{182}  &  4(1)       &  \\
Adder${}_8$            &   900    &             409    &    779    &  429  &        675         &         339       &        606         &         291         &     597     &     295     &     4(1)  & \cellh{514}  & \cellh{256}  &  4(1)       &  \\
RC-Adder${}_6$         &   200    &             93     &    206    &  113  &    393$^\star$     &     164$^\star$   &    \cellh{140}     &     \cellh{71}      &     159     & \cellh{71}  &     1(1)  &     152      &  \cellh{71}  &  1(1)       &  \\
Mod-Red${}_{21}$       &   278    &             105    &    260    &  130  &        217         &         93        &        180         &         77          &     196     &     85      &     3(1)  & \cellh{179}  &  \cellh{76}  &  1(1)       &  \\
Mod-Mult${}_{55}$      &   119    &             48     &    124    &   74  &         91         &         42        &         91         &     \cellh{40}      & \cellh{90}  & \cellh{40}  &     1(1)  &  \cellh{90}  &      41      &  1(1)       &  \\
Toff-Barenco${}_3$     &    58    &             24     &    50     &   26  &     59$^\star$     & \cellh{18}$^\star$&     \cellh{40}     &     \cellh{18}      &     46      &     21      &     1(1)  &  \cellh{40}  &  \cellh{18}  &  3(-5)       &  \\
Toff-NC${}_3$          &    45    &             18     &    41     &   20  &         40         &         16        &     \cellh{35}     &     \cellh{14}      &     36      &     15      &     3(1)  &  \cellh{35}  &  \cellh{14}  &  1(1)       &  \\
Toff-Barenco${}_4$     &   114    &             48     &    117    &   60  &         95         &         44        &     \cellh{72}     &     \cellh{34}      &     88      &     40      &     4(1)  &  \cellh{72}  &  \cellh{34}  &  3(1)       &  \\
Toff-NC${}_4$          &    75    &             30     &    86     &   43  &         65         &         26        &     \cellh{55}     &     \cellh{22}      &     57      &     24      &     3(1)  &  \cellh{55}  &  \cellh{22}  &  1(1)       &  \\
Toff-Barenco${}_5$     &   170    &             72     &    149    &   86  &        140         &         66        &        104         &         50          &     122     &     57      &     4(1)  & \cellh{102}  &  \cellh{48}  &  3(1)       &  \\
Toff-NC${}_5$          &   105    &             42     &    92     &   42  &         90         &         36        &     \cellh{75}     &     \cellh{30}      &     78      &     33      &     3(1)  &  \cellh{75}  &  \cellh{30}  &  1(1)       &  \\
Toff-Barenco${}_{10}$  &   450    &             192    &    392    &  196  &        365         &         176       &    264     &     130     &     325     &     151     &           4(1)     & \cellh{252}  & \cellh{118}  &       3(1)  &  \\
Toff-NC${}_{10}$       &   255    &             102    &    237    &  100  &        215         &         86        &    \cellh{175}     &     \cellh{70}      &     183     &     78      &     3(1) & \cellh{175}  &  \cellh{70}  &  1(1)       &  \\
GF($2^4$)-Mult         &   225    &             99     &    245    &  140  &        193         &         99        &        187         &         99          &     195     &     101     &     2(1) & \cellh{180}  &  \cellh{98}  &  3(-5)       &  \\
GF($2^5$)-Mult         &   347    &         \cellh{154}&    351    &  197  &        304         &     \cellh{154}   &        296         &     \cellh{154}     &     306     &     156     &     1(1) & \cellh{289}  &     155      &  4(-20)       &  \\
GF($2^6$)-Mult         &   495    &             221    &    545    &  308  &        422         &         221       &        403         &         221         &     418     & \cellh{217} &     4(-5) & \cellh{390}  &     218      &  3(-5)       &  \\
GF($2^7$)-Mult         &   669    &             300    &    736    &  417  &        573         &         300       &        555         &         300         &     572     &     299     &     4(-5) & \cellh{535}  & \cellh{292}  &  4(-20)       &  \\
GF($2^8$)-Mult         &   883    &             405    &   1015    &  606  &        745         &         405       &        712         &         405         &     745     &     405     &     1(1) & \cellh{691}  & \cellh{399}  &  1(1)       &  \\ \midrule
Avg. reduction         &          &   & $\sim3\%$ &         \cellb{$\sim-22\%$}          &     $\sim14\%$     &           $\sim9\%$           & $\sim27\%$ & $\sim19\%$  & $\sim23\%$  & $\sim16\%$  & &  \cellh{$\sim29\%$}  &  \cellh{$\sim21\%$}  & &  \\ \bottomrule
	\end{tabular}

%% file: tabulars/tpar_todd.tex
\resizebox{.485\textwidth}{!}{%
\setlength{\tabcolsep}{0.6mm}
\centering
\begin{tabular}{l|c|c|c|c|c|c|c|c|c|c}
	\toprule
	\rule{0pt}{1\normalbaselineskip} \multirow{2}{1pt}{\textbf{Circuit}} & \multicolumn{3}{c|}{\textbf{Original}} & \multicolumn{3}{c|}{\textbf{PyZX+TODD}} & \multicolumn{4}{c}{\textbf{Heuristic+TODD}} \\ \cline{2-11}
	\rule{0pt}{1\normalbaselineskip}                                     &  $\Sigma$   &    $2Q$     &        $\mathcal{T}$         & $\Sigma$ & $2Q$ &     $\mathcal{T}$     & $\Sigma$ & $2Q$ & $\mathcal{T}$ &   Alg.    \\ \midrule
	CSLA-MUX${}_3$                  & 170 & 80  &              70              &   262    & 175  &      43       &   257    & 169  &  43   &     2(1)     \\
	CSUM-MUX${}_3$                                                       & 420 & 168 &              196              &   575    & 428  &      74       &   411    & 261  &  74   &     4(-5)     \\
	QCLA-Com${}_7$                                                       & 443 & 186 &              203              &   454    & 274  &      93       &   389    & 211  &  93   &     4(1)     \\ 
	QCLA-Adder${}_{10}$                                                  & 521 & 233 &             238              &   800    & 517  &      143      &   677    & 391  &  143  &     4(-20)     \\ 
	Mod-Mult${}_{55}$                                                    & 119  & 48  &              49              &   107    &  56  &      27       &   104    &  55  &  27   &     4(-5)     \\ 
	GF($2^4$)-Mult                                                       & 225 & 99  &              112              &   298    & 221  &      52       &   295    & 220  &  52   &     1(-5)     \\
	GF($2^5$)-Mult                                                       & 347 & 154 &             175              &   538    & 420  &      88       &   524    & 403  &  88   &     3(1)     \\
	GF($2^6$)-Mult                                                       & 495 & 221 &             252              &   943    & 764  &      134      &   933    & 750  &  134  &     3(1)     \\
	GF($2^7$)-Mult                                                       & 669 & 300 &             343              &   1253   & 1036 &      180      &   1223   & 993  &  180  &     4(1)     \\
	GF($2^8$)-Mult                                                       & 883 & 405 &             448              &   1791   & 1521 &      224      &   1780   & 1507 &  224  &     3(1)     \\ \bottomrule
\end{tabular}
}

%% file: content/conclusion.tex
\section{Conclusions and Future Work}%
\label{sec:conclusion}
In this work we introduce two functions, namely the Local Complementation Heuristic LCH (for the local complementation rule) and the Pivot Heuristic PH (for the pivot rule).
The functions calculate the number of Hadamard wires that would be added or removed by applying the respective rule, thus serving as a heuristic for estimating the 2-qubit gate count of the underlying circuit.
This allows us to develop a more sophisticated strategy for ZX-diagram simplification: First, dismiss the applicable rules that cost too much and then either select a rule randomly or select the rule with the best wire count decrease.

Notably, the T gate count remains unchanged throughout this process, which is why our approach and others that mainly decrease the T gate count complement each other well.  
Further, we introduce the new \textit{Neighbor Unfusion} rule which combines the established \textit{fusion} and \textit{identity} rules. This rule allows introducing spiders with arbitrary phases into the circuit if needed, for example when the local complementation or pivot rule would be useful to reduce Hadamard wires. 
As a side note, we also formally describe how to use the local complementation and the pivoting rule on spiders with non-Clifford phases, which is a common implementation practice but has never been mentioned in theory.

We measure the impact of aligning the optimization strategy with the heuristics and adding the neighbor unfusion rule by comparing our algorithm to four other approaches, some based on ZX-calculus and some not, on a set of 24 well-established benchmark circuits. 
Our approaches show significant improvements compared to all other ZX-based approaches, especially in 2-qubit gate reduction. 
On their own, non-ZX-based approaches still yield slightly better results than our ZX-based approaches. However, when combining both we are able optimize circuits better than the previously best known result, which seems to be a promising field for further research.

Using heuristics for ZX-diagram simplification also provides many possibilities for future improvement. 
Regarding the selection of rules, both random and greedy strategy are non-optimal for finding a ZX-diagram with minimal number of wires. 
Instead, we propose using a metaheuristic selection strategy like simulated annealing for escaping local minima during simplification. 
Furthermore, since simplification with neighbor unfusion tends to yield the best results, we think it is important to further investigate in which cases neighbor unfusion generates $XY$ spiders and if we can preserve valid ZX-diagrams when allowing unfusion on spiders which get extracted on different qubits.

%% file: content/acks.tex
\section*{Acknowledgment}
This work is partially supported by the German Federal Ministry of Education and Research (BMBF) under the funding programme Quantum Technologies - From Basic Research to Market under contract number 13N16077.

%% file: content/appendix-further-rules.tex
\section{Further rules}\label{app:further-rules}
In addition to the rules in \autoref{sec:related}, additional rules have been developed to eliminate \textit{every} interior Clifford spider.

\subsection{Pivoting Boundary Spiders (p1)}
The pivoting rule can also be applied if one of the spiders is a boundary spider, i.e., connected to an input or output, using the following transformation: 
\begin{equation}
	\input{pivotBoundaryRule.tikz}
\end{equation}
Here $v$ gets transformed to an interior spider and both $u$ and $v$ can be removed using the pivoting rule.

\begin{wrapfigure}{r}{0.4\textwidth}
	\vspace{-3em}
	\begin{equation}
		\input{gadget_fuse_rule.tikz}
	\end{equation}
	\vspace{-1.75em}
\end{wrapfigure}
\subsection{Gadget Fusion (gf):}
An important feature of phase gadgets is that we can fuse two phase gadgets connected to the same neighbors by summing up their phases.

This rule is used for eliminating non-Clifford spiders in a diagram, for instance, two phase gadgets with phase $\pi/4$ connected to the same set of neighbours can be fused into a single phase gadget with phase $\pi/2$.
Combining the Clifford simplification algorithm with those extended rules we can eliminate \textit{all} interior Clifford spiders (in exchange for phase gadgets) and \textit{some} interior non-Clifford spiders.

%% file: tikz/pivotBoundaryRule.tikz
\begin{tikzpicture}
	\begin{pgfonlayer}{nodelayer}
		\node [style=Z phase dot] (0) at (-6, 0) {$j\pi$};
		\node [style=none] (1) at (-7, -2) {};
		\node [style=none] (2) at (-5, -2) {};
		\node [style=Z phase dot] (3) at (-3, 0) {$k\pi$};
		\node [style=none] (4) at (-4, -2) {};
		\node [style=none] (5) at (-2, -2) {};
		\node [style=none] (6) at (-1.5, 0) {};
		\node [style=none] (7) at (-0.25, -0.25) {$=$};
		\node [style=none] (8) at (-0.25, 0.5) {$(i1,i2)$};
		\node [style=Z phase dot] (9) at (1.75, 0) {$j\pi$};
		\node [style=none] (10) at (0.75, -2) {};
		\node [style=none] (11) at (2.75, -2) {};
		\node [style=Z phase dot] (12) at (4.75, 0) {$k\pi$};
		\node [style=none] (13) at (3.75, -2) {};
		\node [style=none] (14) at (5.75, -2) {};
		\node [style=none] (15) at (7.5, 0) {};
		\node [style=Z dot] (16) at (6.25, 0) {};
		\node [style=none] (17) at (-6, -1.75) {$\ldots$};
		\node [style=none] (18) at (-3, -1.75) {$\ldots$};
		\node [style=none] (19) at (1.725, -1.75) {$\ldots$};
		\node [style=none] (20) at (4.75, -1.75) {$\ldots$};
		\node [style=none] (21) at (-6, 0.75) {$u$};
		\node [style=none] (22) at (-3, 0.75) {$v$};
		\node [style=none] (23) at (1.75, 0.75) {$u$};
		\node [style=none] (24) at (4.75, 0.75) {$v$};
		\node [style=none] (26) at (8.5, -2) {};
		\node [style=none] (27) at (10.5, -2) {};
		\node [style=none] (29) at (11.5, -2) {};
		\node [style=none] (30) at (13.5, -2) {};
		\node [style=none] (31) at (12.25, 0) {};
		\node [style=Z dot] (32) at (11, 0) {};
		\node [style=none] (33) at (9.525, -1.75) {$\ldots$};
		\node [style=none] (34) at (12.3, -1.75) {$\ldots$};
		\node [style=none] (35) at (9, -0.25) {$=$};
		\node [style=none] (36) at (9, 0.5) {$(p)$};
		\node [style=none] (37) at (-2.75, 0.5) {};
		\node [style=none] (38) at (-2.75, -0.5) {};
		\node [style=none] (39) at (-1.5, 0.5) {};
		\node [style=none] (40) at (-1.5, -0.5) {};
		\node [style=none] (41) at (1.25, 1) {};
		\node [style=none] (42) at (1.25, -0.5) {};
		\node [style=none] (43) at (5.5, 1) {};
		\node [style=none] (44) at (5.5, -0.5) {};
	\end{pgfonlayer}
	\begin{pgfonlayer}{edgelayer}
		\draw [style=hadamard edge] (0) to (1.center);
		\draw [style=hadamard edge] (0) to (2.center);
		\draw [style=hadamard edge] (0) to (3);
		\draw [style=hadamard edge] (3) to (4.center);
		\draw [style=hadamard edge] (3) to (5.center);
		\draw (3) to (6.center);
		\draw [style=hadamard edge] (9) to (10.center);
		\draw [style=hadamard edge] (9) to (11.center);
		\draw [style=hadamard edge] (9) to (12);
		\draw [style=hadamard edge] (12) to (13.center);
		\draw [style=hadamard edge] (12) to (14.center);
		\draw [style=hadamard edge] (12) to (16);
		\draw [style=hadamard edge] (16) to (15.center);
		\draw [style=hadamard edge] (32) to (31.center);
		\draw [style=hadamard edge] (32) to (26.center);
		\draw [style=hadamard edge] (32) to (27.center);
		\draw [style=hadamard edge] (32) to (29.center);
		\draw [style=hadamard edge] (32) to (30.center);
		\draw [style=gray edge] (37.center) to (39.center);
		\draw [style=gray edge] (39.center) to (40.center);
		\draw [style=gray edge] (40.center) to (38.center);
		\draw [style=gray edge] (38.center) to (37.center);
		\draw [style=gray edge] (41.center) to (43.center);
		\draw [style=gray edge] (43.center) to (44.center);
		\draw [style=gray edge] (44.center) to (42.center);
		\draw [style=gray edge] (42.center) to (41.center);
	\end{pgfonlayer}
\end{tikzpicture}

%% file: tikz/gadget_fuse_rule.tikz
\begin{tikzpicture}
	\begin{pgfonlayer}{nodelayer}
		\node [style=Z phase dot] (0) at (-3, 0.75) {$\alpha$};
		\node [style=Z phase dot] (1) at (-3, -0.75) {$\beta$};
		\node [style=Z phase dot] (2) at (0, 1.25) {$\alpha_1$};
		\node [style=Z phase dot] (3) at (0, -1.25) {$\alpha_n$};
		\node [style=Z phase dot] (4) at (3.5, 0) {$\alpha+\beta$};
		\node [style=none] (11) at (0, 0) {$\vdots$};
		\node [style=Z phase dot] (12) at (6.5, 1.25) {$\alpha_1$};
		\node [style=Z phase dot] (13) at (6.5, -1.25) {$\alpha_n$};
		\node [style=none] (20) at (6.5, 0) {$\vdots$};
		\node [style=Z dot] (21) at (5, 0) {};
		\node [style=Z dot] (22) at (-2, 0.75) {};
		\node [style=Z dot] (23) at (-2, -0.75) {};
		\node [style=none] (24) at (1.75, 0) {$\overset{(gf)}{=}$};
		\node [style=none] (26) at (-3.5, 1.25) {};
		\node [style=none] (27) at (-3.5, -1.25) {};
		\node [style=none] (28) at (-1.75, 1.25) {};
		\node [style=none] (29) at (-1.75, -1.25) {};
	\end{pgfonlayer}
	\begin{pgfonlayer}{edgelayer}
		\draw [style=hadamard edge] (4) to (21);
		\draw [style=hadamard edge] (21) to (12);
		\draw [style=hadamard edge] (21) to (13);
		\draw [style=hadamard edge] (0) to (22);
		\draw [style=hadamard edge] (22) to (3);
		\draw [style=hadamard edge] (22) to (2);
		\draw [style=hadamard edge] (2) to (23);
		\draw [style=hadamard edge] (23) to (3);
		\draw [style=hadamard edge] (23) to (1);
		\draw [style=gray edge] (26.center) to (28.center);
		\draw [style=gray edge] (28.center) to (29.center);
		\draw [style=gray edge] (29.center) to (27.center);
		\draw [style=gray edge] (27.center) to (26.center);
	\end{pgfonlayer}
\end{tikzpicture}

%% file: content/appendix-graph-theory.tex
\section{Graph Theory}\label{sec:background-graph}
Since ZX-calculus and its optimization strategies rely on graph operations on undirected graphs, we provide some background on it:
An \textit{undirected graph} is a tuple $G=(V,E)$ with \textit{vertices} (or ``nodes'') $V$ and \textit{edges} $E \subseteq V\times V$.

\subsection{Local Complementation} \label{sec:background-graph-lc} 
The local complementation~$\star$~\cite{Bouchet.1988} of an undirected graph $G=\left(V,E\right)$ about a vertex~$u$ is defined as follows ($\Delta$ is the symmetric set difference: $  A \mathrel{\Delta} B \coloneqq \left(A\cup B\right)\backslash \left(A\cap B\right)$):
\begin{equation}
	G\star u \coloneqq \left(V,E \mathrel{\Delta} \left\{\left(a,b\right)| \left(a,u\right),\left(b,u\right) \in E, a\neq b\right\}\right)
\end{equation}

The following example shows a graph $ G $ and its local complementation about $ a $. Intuitively, local complementation connects two neighbours of $ a $ if they are \textit{not} connected (e.g.,~$ b,c $) and disconnects them otherwise (e.g.,~$ c,d $).
\begin{equation}\label{eq:lcompGraph}
	\input{lcompGraph.tikz}
\end{equation}

\subsection{Pivoting~\cite{Geelen.2009}} \label{sec:background-graph-pv} 
Pivoting~$ \land $ rewrites an edge $(u,v) \in E$ by triple local complementation:
\begin{equation}
	G\land uv \coloneqq ((G\star u)\star v)\star u
\end{equation}
To derive the new graph, we consider three disjoint sets (where the neighborhood of vertex $ x $ is defined as   $ N(x) = \{y \in V|(x,y)\in E\} $):
	\begin{itemize}
		\item $A \coloneqq N(u) \cap N(v)$: Vertices connected to $u$ and $v$.
		\item $B \coloneqq N(u) \backslash N(v)$: Vertices connected to $u$ and not to $v$.
		\item $C \coloneqq N(v) \backslash N(u)$: Vertices connected to $v$ and not to $u$.
	\end{itemize}

\noindent
In a pivoted graph $G\land uv$, two vertices from different sets $ A,B$ or $ C $ are connected if, and only if, the two are not connected in~$G$. Connections between vertices of the same set are not modified.
As an example, consider the following graphs~$ G $ (left) and $ G\land uv $ (right), where $ A=\{b\}, B=\{a,d\}, C = \{c,e\}$. Intuitively, pivoting connects all vertices between $ A,B,C $ that are not connected in $ G $ (e.g.,~$a,b$) and disconnects them otherwise (e.g.,~$b,d$):
\begin{equation}
	\input{pivotGraph.tikz}
\end{equation}

%% file: tikz/lcompGraph.tikz
\begin{tikzpicture}
	\begin{pgfonlayer}{nodelayer}
		\node [style=vertex] (0) at (-4.75, 2) {};
		\node [style=vertex] (1) at (-3.5, 2) {};
		\node [style=vertex] (2) at (-3.5, 0.75) {};
		\node [style=vertex] (3) at (-4.75, 0.75) {};
		\node [style=none] (4) at (-5.5, 1.5) {$G=$};
		\node [style=none] (5) at (0.25, 1.5) {$(G\star a)=$};
		\node [style=none] (6) at (-4.75, 2.75) {$a$};
		\node [style=none] (7) at (-3.25, 2.75) {$b$};
		\node [style=none] (8) at (-4.75, 0) {$c$};
		\node [style=none] (9) at (-3.25, 0) {$d$};
		\node [style=vertex] (10) at (1.75, 2) {};
		\node [style=vertex] (11) at (3, 2) {};
		\node [style=vertex] (12) at (3, 0.75) {};
		\node [style=vertex] (13) at (1.75, 0.75) {};
		\node [style=none] (14) at (1.75, 2.75) {$a$};
		\node [style=none] (15) at (3, 2.75) {$b$};
		\node [style=none] (16) at (1.75, 0) {$c$};
		\node [style=none] (17) at (3, 0) {$d$};
	\end{pgfonlayer}
	\begin{pgfonlayer}{edgelayer}
		\draw (0) to (1);
		\draw (0) to (3);
		\draw (0) to (2);
		\draw (10) to (11);
		\draw (10) to (13);
		\draw (10) to (12);
		\draw (13) to (11);
		\draw (11) to (12);
		\draw (2) to (3);
	\end{pgfonlayer}
\end{tikzpicture}

%% file: tikz/pivotGraph.tikz
\begin{tikzpicture}
	\begin{pgfonlayer}{nodelayer}
		\node [style=vertex] (0) at (-3, 2.5) {};
		\node [style=vertex] (1) at (-0.25, 2.5) {};
		\node [style=vertex] (2) at (0.5, 1) {};
		\node [style=vertex] (3) at (-1.75, 0.75) {};
		\node [style=vertex] (4) at (0.5, -0.25) {};
		\node [style=vertex] (5) at (-4, 1) {};
		\node [style=vertex] (6) at (-4, -0.25) {};
		\node [style=none] (7) at (-4.75, 1) {$a$};
		\node [style=none] (8) at (-4.75, -0.25) {$d$};
		\node [style=none] (9) at (-3, 3) {$u$};
		\node [style=none] (10) at (-0.25, 3) {$v$};
		\node [style=none] (11) at (1.25, 1) {$c$};
		\node [style=none] (12) at (1.25, -0.25) {$e$};
		\node [style=none] (13) at (-1.75, 1.5) {$b$};
		\node [style=none] (29) at (2.5, 1.75) {$\xrightarrow{G \land uv}$};
		\node [style=vertex] (30) at (5.25, 2.5) {};
		\node [style=vertex] (31) at (8, 2.5) {};
		\node [style=vertex] (32) at (8.75, 1) {};
		\node [style=vertex] (33) at (6.5, 0.75) {};
		\node [style=vertex] (34) at (8.75, -0.25) {};
		\node [style=vertex] (35) at (4.25, 1) {};
		\node [style=vertex] (36) at (4.25, -0.25) {};
		\node [style=none] (37) at (3.5, 1) {$a$};
		\node [style=none] (38) at (3.5, -0.25) {$d$};
		\node [style=none] (39) at (5.25, 3) {$v$};
		\node [style=none] (40) at (8, 3) {$u$};
		\node [style=none] (41) at (9.5, 1) {$c$};
		\node [style=none] (42) at (9.5, -0.25) {$e$};
		\node [style=none] (43) at (6.5, 1.5) {$b$};
	\end{pgfonlayer}
	\begin{pgfonlayer}{edgelayer}
		\draw (0) to (1);
		\draw (1) to (2);
		\draw (1) to (4);
		\draw (1) to (3);
		\draw (3) to (0);
		\draw (0) to (5);
		\draw (6) to (0);
		\draw (30) to (31);
		\draw (31) to (32);
		\draw (31) to (34);
		\draw (31) to (33);
		\draw (33) to (30);
		\draw (30) to (35);
		\draw (36) to (30);
		\draw (36) to (32);
		\draw (35) to (32);
		\draw (4) to (3);
		\draw (3) to (6);
		\draw (35) to (33);
		\draw (33) to (32);
		\draw (36) to (34);
		\draw (34) to (35);
	\end{pgfonlayer}
\end{tikzpicture}

%% file: content/appendix-zx-example.tex
\section{Example for ZX optimization}\label{app:zx-example}
The following circuit can be optimized as follows:
\begin{equation}
\begin{array}{c}
\Qcircuit @C=1em @R=.5em @!R {
		& \ctrl{2} & \gate{Z} & \ctrl{1} & \gate{Z} & \qw \\
		& \qw      & \ctrl{1} & \targ    & \gate{X} & \qw \\
		& \targ    & \targ    & \qw      & \gate{X} & \qw
	} \end{array}
=
\begin{array}{c}
\Qcircuit @C=1em @R=.5em @!R {
	& \ctrl{1} & \qw      & \qw      & \qw \\
	& \targ    & \gate{X} & \ctrl{1} & \qw \\
	& \qw      & \qw      & \targ    & \qw
}
\end{array}
\end{equation}
We use the following rules\footnote{The example is inspired by a talk of Russ Duncan ``Quantum Formal Methods'' from 2021 (1h 35min) which is publicly available.} (the affected spiders/wires to which a rule is applied are framed):\\

\begin{tabularx}{.9\textwidth}{>{\linewidth=0.35\textwidth}X>{\linewidth=0.5\textwidth}c}
	\begin{enumerate}[label=\arabic*),leftmargin=.5em]
		\vspace{-1em}
		\item Eliminate the two Z-gates using spider fusion (f): 
	\end{enumerate}&
	$ \input{example_circuit_1.tikz}  \xrightarrow{(\textbf{\textit{f}})} \input{example_circuit_3.tikz} $\\
	\begin{enumerate}[label=\arabic*),start=2,leftmargin=.5em]	
		\vspace{-1.5em}
		\item Reduce from 3 to 2 CNOTs with fusion (f) and bialgebra rule (b): 
	\end{enumerate}&
	$ \xrightarrow{(\textbf{\textit{f}})}  \input{example_circuit_4.tikz}  \xrightarrow{(\textbf{\textit{f}})}  \input{example_circuit_5.tikz} \xrightarrow{(\textbf{\textit{b}})}  \input{example_circuit_8.tikz} $ \\
	\begin{enumerate}[label=\arabic*),start=3,leftmargin=.5em]	
		\vspace{-1em}
		\item Eliminate one X by the $ \pi $ copy rule:
	\end{enumerate}&
	$  \xrightarrow{(\mathbf{\pi})}  \input{example_circuit_9.tikz} \xrightarrow{(\textbf{\textit{f}})}  \input{example_circuit_10.tikz} $
\end{tabularx}

%% file: tikz/example_circuit_1.tikz
\begin{tikzpicture}
	\begin{pgfonlayer}{nodelayer}
		\node [style=none] (0) at (6, 1.25) {};
		\node [style=none] (1) at (6, 0.25) {};
		\node [style=none] (2) at (6, -0.75) {};
		\node [style=X dot] (5) at (6.75, -0.75) {};
		\node [style=X dot] (7) at (7.75, -0.75) {};
		\node [style=X dot] (12) at (9.75, 0.25) {$\pi$};
		\node [style=none] (14) at (10.5, 0.25) {};
		\node [style=none] (15) at (10.5, 1.25) {};
		\node [style=Z dot] (17) at (7.75, 1.25) {$\pi$};
		\node [style=Z dot] (18) at (8.75, 1.25) {};
		\node [style=X dot] (19) at (8.75, 0.25) {};
		\node [style=Z dot] (22) at (7.75, 0.25) {};
		\node [style=Z dot] (23) at (6.75, 1.25) {};
		\node [style=none] (26) at (10.5, -0.75) {};
		\node [style=none] (27) at (9.75, 1.25) {};
		\node [style=Z dot] (28) at (9.75, 1.25) {$\pi$};
		\node [style=X dot] (29) at (9.75, -0.75) {$\pi$};
		\node [style=none] (34) at (7.25, 1.75) {};
		\node [style=none] (35) at (7.25, 0.75) {};
		\node [style=none] (36) at (10.25, 1.75) {};
		\node [style=none] (37) at (10.25, 0.75) {};
	\end{pgfonlayer}
	\begin{pgfonlayer}{edgelayer}
		\draw (0.center) to (23);
		\draw (18) to (17);
		\draw (12) to (19);
		\draw (23) to (5);
		\draw (22) to (7);
		\draw (18) to (19);
		\draw (1.center) to (22);
		\draw (2.center) to (5);
		\draw (5) to (7);
		\draw (22) to (19);
		\draw (18) to (28);
		\draw (17) to (23);
		\draw (12) to (14.center);
		\draw (28) to (15.center);
		\draw (7) to (29);
		\draw (29) to (26.center);
		\draw [style=gray edge] (34.center) to (36.center);
		\draw [style=gray edge] (36.center) to (37.center);
		\draw [style=gray edge] (37.center) to (35.center);
		\draw [style=gray edge] (35.center) to (34.center);
	\end{pgfonlayer}
\end{tikzpicture}

%% file: tikz/example_circuit_3.tikz
\begin{tikzpicture}
	\begin{pgfonlayer}{nodelayer}
		\node [style=none] (0) at (5.75, 1.25) {};
		\node [style=none] (1) at (5.75, 0.25) {};
		\node [style=none] (2) at (5.75, -0.75) {};
		\node [style=X dot] (5) at (6.5, -0.75) {};
		\node [style=X dot] (7) at (7.5, -0.75) {};
		\node [style=X dot] (12) at (9.5, 0.25) {$\pi$};
		\node [style=none] (14) at (10.25, 0.25) {};
		\node [style=none] (15) at (10.25, 1.25) {};
		\node [style=Z dot] (18) at (8.5, 1.25) {};
		\node [style=X dot] (19) at (8.5, 0.25) {};
		\node [style=Z dot] (22) at (7.5, 0.25) {};
		\node [style=Z dot] (23) at (6.5, 1.25) {};
		\node [style=none] (26) at (10.25, -0.75) {};
		\node [style=X dot] (29) at (9.5, -0.75) {$\pi$};
		\node [style=none] (30) at (6.25, 1.75) {};
		\node [style=none] (31) at (6.25, 0.75) {};
		\node [style=none] (32) at (8.75, 1.75) {};
		\node [style=none] (33) at (8.75, 0.75) {};
		\node [style=none] (34) at (6.25, -0.25) {};
		\node [style=none] (35) at (6.25, -1.25) {};
		\node [style=none] (36) at (7.75, -0.25) {};
		\node [style=none] (37) at (7.75, -1.25) {};
	\end{pgfonlayer}
	\begin{pgfonlayer}{edgelayer}
		\draw (0.center) to (23);
		\draw (12) to (19);
		\draw (23) to (5);
		\draw (22) to (7);
		\draw (18) to (19);
		\draw (1.center) to (22);
		\draw (2.center) to (5);
		\draw (5) to (7);
		\draw (22) to (19);
		\draw (12) to (14.center);
		\draw (23) to (18);
		\draw (7) to (29);
		\draw (29) to (26.center);
		\draw (18) to (15.center);
		\draw [style=gray edge] (30.center) to (32.center);
		\draw [style=gray edge] (32.center) to (33.center);
		\draw [style=gray edge] (33.center) to (31.center);
		\draw [style=gray edge] (31.center) to (30.center);
		\draw [style=gray edge] (34.center) to (36.center);
		\draw [style=gray edge] (36.center) to (37.center);
		\draw [style=gray edge] (37.center) to (35.center);
		\draw [style=gray edge] (35.center) to (34.center);
	\end{pgfonlayer}
\end{tikzpicture}

%% file: tikz/example_circuit_4.tikz
\begin{tikzpicture}
	\begin{pgfonlayer}{nodelayer}
		\node [style=none] (0) at (6, 1.25) {};
		\node [style=none] (1) at (6, 0.25) {};
		\node [style=none] (2) at (6, -0.75) {};
		\node [style=X dot] (5) at (7.75, -0.75) {};
		\node [style=none] (14) at (9.25, 0.25) {};
		\node [style=none] (15) at (9.25, 1.25) {};
		\node [style=X dot] (19) at (7.75, 0.25) {};
		\node [style=Z dot] (22) at (6.75, 0.25) {};
		\node [style=Z dot] (23) at (6.75, 1.25) {};
		\node [style=none] (26) at (9.25, -0.75) {};
		\node [style=X dot] (27) at (8.5, 0.25) {$\pi$};
		\node [style=X dot] (28) at (8.5, -0.75) {$\pi$};
		\node [style=none] (29) at (6.4, 1.75) {};
		\node [style=none] (30) at (6.4, 0.75) {};
		\node [style=none] (31) at (7.15, 1.75) {};
		\node [style=none] (32) at (7.15, 0.75) {};
		\node [style=none] (33) at (7.35, -0.25) {};
		\node [style=none] (34) at (7.35, -1.25) {};
		\node [style=none] (35) at (8.1, -0.25) {};
		\node [style=none] (36) at (8.1, -1.25) {};
	\end{pgfonlayer}
	\begin{pgfonlayer}{edgelayer}
		\draw (23) to (5);
		\draw (1.center) to (22);
		\draw (2.center) to (5);
		\draw (22) to (19);
		\draw (19) to (14.center);
		\draw (23) to (15.center);
		\draw (23) to (19);
		\draw (5) to (22);
		\draw (0.center) to (23);
		\draw (5) to (28);
		\draw (28) to (26.center);
		\draw [style=gray edge] (29.center) to (31.center);
		\draw [style=gray edge] (31.center) to (32.center);
		\draw [style=gray edge] (32.center) to (30.center);
		\draw [style=gray edge] (30.center) to (29.center);
		\draw [style=gray edge] (33.center) to (35.center);
		\draw [style=gray edge] (35.center) to (36.center);
		\draw [style=gray edge] (36.center) to (34.center);
		\draw [style=gray edge] (34.center) to (33.center);
	\end{pgfonlayer}
\end{tikzpicture}

%% file: tikz/example_circuit_5.tikz
\begin{tikzpicture}
	\begin{pgfonlayer}{nodelayer}
		\node [style=none] (0) at (-1.75, 1.75) {};
		\node [style=none] (1) at (-1.75, 0.25) {};
		\node [style=none] (2) at (-1.75, -1.25) {};
		\node [style=X dot] (3) at (0, -0.5) {};
		\node [style=none] (4) at (1.5, 0.25) {};
		\node [style=none] (5) at (1.5, 1.75) {};
		\node [style=X dot] (6) at (0, 0.25) {};
		\node [style=Z dot] (7) at (-1, 0.25) {};
		\node [style=Z dot] (8) at (-1, 1) {};
		\node [style=none] (9) at (1.5, -1.25) {};
		\node [style=X dot] (10) at (0.75, 0.25) {$\pi$};
		\node [style=X dot] (11) at (0.75, -1.25) {$\pi$};
		\node [style=Z dot] (12) at (-1, 1.75) {};
		\node [style=X dot] (13) at (0, -1.25) {};
		\node [style=none] (14) at (-1.25, 1.25) {};
		\node [style=none] (15) at (-1.25, -0.75) {};
		\node [style=none] (16) at (0.25, 1.25) {};
		\node [style=none] (17) at (0.25, -0.75) {};
	\end{pgfonlayer}
	\begin{pgfonlayer}{edgelayer}
		\draw (8) to (3);
		\draw (1.center) to (7);
		\draw (7) to (6);
		\draw (6) to (4.center);
		\draw (8) to (6);
		\draw (3) to (7);
		\draw (11) to (9.center);
		\draw (0.center) to (12);
		\draw (12) to (5.center);
		\draw (12) to (8);
		\draw (13) to (11);
		\draw (13) to (3);
		\draw (13) to (2.center);
		\draw [style=gray edge] (14.center) to (16.center);
		\draw [style=gray edge] (16.center) to (17.center);
		\draw [style=gray edge] (17.center) to (15.center);
		\draw [style=gray edge] (15.center) to (14.center);
	\end{pgfonlayer}
\end{tikzpicture}

%% file: tikz/example_circuit_8.tikz
\begin{tikzpicture}
	\begin{pgfonlayer}{nodelayer}
		\node [style=none] (0) at (7, 1.25) {};
		\node [style=none] (1) at (7, 0.25) {};
		\node [style=none] (2) at (7, -0.75) {};
		\node [style=none] (14) at (10, 0.25) {};
		\node [style=none] (15) at (10, 1.25) {};
		\node [style=none] (26) at (10, -0.75) {};
		\node [style=Z dot] (27) at (7.75, 1.25) {};
		\node [style=Z dot] (28) at (8.5, 0.25) {};
		\node [style=X dot] (29) at (9.25, 0.25) {$\pi$};
		\node [style=X dot] (30) at (9.25, -0.75) {$\pi$};
		\node [style=X dot] (31) at (8.5, -0.75) {};
		\node [style=X dot] (32) at (7.75, 0.25) {};
		\node [style=none] (33) at (8, 0.75) {};
		\node [style=none] (34) at (8, -0.25) {};
		\node [style=none] (35) at (9.75, 0.75) {};
		\node [style=none] (36) at (9.75, -0.25) {};
	\end{pgfonlayer}
	\begin{pgfonlayer}{edgelayer}
		\draw (27) to (0.center);
		\draw (27) to (15.center);
		\draw (30) to (26.center);
		\draw (29) to (14.center);
		\draw (29) to (28);
		\draw (31) to (30);
		\draw (31) to (2.center);
		\draw (31) to (28);
		\draw (1.center) to (32);
		\draw (32) to (28);
		\draw (32) to (27);
		\draw [style=gray edge] (33.center) to (35.center);
		\draw [style=gray edge] (35.center) to (36.center);
		\draw [style=gray edge] (36.center) to (34.center);
		\draw [style=gray edge] (34.center) to (33.center);
	\end{pgfonlayer}
\end{tikzpicture}

%% file: tikz/example_circuit_9.tikz
\begin{tikzpicture}
	\begin{pgfonlayer}{nodelayer}
		\node [style=none] (0) at (7, 1.25) {};
		\node [style=none] (1) at (7, 0.25) {};
		\node [style=none] (2) at (7, -1.25) {};
		\node [style=none] (14) at (11.25, 0.25) {};
		\node [style=none] (15) at (11.25, 1.25) {};
		\node [style=none] (26) at (11.25, -1.25) {};
		\node [style=Z dot] (27) at (7.75, 1.25) {};
		\node [style=Z dot] (28) at (9.25, 0.25) {};
		\node [style=X dot] (31) at (9.25, -1.25) {};
		\node [style=X dot] (32) at (7.75, 0.25) {};
		\node [style=X dot] (33) at (8.5, 0.25) {$\pi$};
		\node [style=X dot] (34) at (9.25, -0.5) {$\pi$};
		\node [style=X dot] (35) at (10.25, -1.25) {$\pi$};
		\node [style=none] (36) at (8.9, -0.1) {};
		\node [style=none] (37) at (8.9, -1.6) {};
		\node [style=none] (38) at (10.65, -0.1) {};
		\node [style=none] (39) at (10.65, -1.6) {};
	\end{pgfonlayer}
	\begin{pgfonlayer}{edgelayer}
		\draw (27) to (0.center);
		\draw (27) to (15.center);
		\draw (31) to (2.center);
		\draw (1.center) to (32);
		\draw (32) to (27);
		\draw (32) to (33);
		\draw (33) to (28);
		\draw (28) to (14.center);
		\draw (28) to (34);
		\draw (34) to (31);
		\draw (31) to (35);
		\draw (35) to (26.center);
		\draw [style=gray edge] (36.center) to (38.center);
		\draw [style=gray edge] (38.center) to (39.center);
		\draw [style=gray edge] (39.center) to (37.center);
		\draw [style=gray edge] (37.center) to (36.center);
	\end{pgfonlayer}
\end{tikzpicture}

%% file: tikz/example_circuit_10.tikz
\begin{tikzpicture}
	\begin{pgfonlayer}{nodelayer}
		\node [style=none] (0) at (7, 1.25) {};
		\node [style=none] (1) at (7, 0.25) {};
		\node [style=none] (2) at (7, -0.75) {};
		\node [style=none] (14) at (10.25, 0.25) {};
		\node [style=none] (15) at (10.25, 1.25) {};
		\node [style=none] (26) at (10.25, -0.75) {};
		\node [style=Z dot] (27) at (7.75, 1.25) {};
		\node [style=Z dot] (28) at (9.25, 0.25) {};
		\node [style=X dot] (31) at (9.25, -0.75) {};
		\node [style=X dot] (32) at (7.75, 0.25) {};
		\node [style=X dot] (33) at (8.5, 0.25) {$\pi$};
	\end{pgfonlayer}
	\begin{pgfonlayer}{edgelayer}
		\draw (27) to (0.center);
		\draw (27) to (15.center);
		\draw (31) to (2.center);
		\draw (1.center) to (32);
		\draw (32) to (27);
		\draw (32) to (33);
		\draw (33) to (28);
		\draw (28) to (14.center);
		\draw (28) to (31);
		\draw (31) to (26.center);
	\end{pgfonlayer}
\end{tikzpicture}